%% file: main_ndss2022.tex
\documentclass[conference]{IEEEtran} 
\include{moritastyle}

\include{preamble}

\usepackage{tikz,mathrsfs}
\usepackage{multirow}

\let\labelindent\relax 
\usepackage{enumitem}
\usepackage{booktabs}
\usepackage{threeparttable}

\newcommand{\jacob}[1]{}
\newcommand{\nuts}[1]{}
\newcommand{\matsuda}[1]{}
\newcommand{\rnote}[1]{}

\newcommand{\pre}[1]{\langle #1 \rangle}

\newcommand{\ignore}[1]{}

\newcommand{\sett}[1]{\ensuremath{\mathbb{#1}}}

\newcommand{\Fset}{\ensuremath{\sett{F}}}

\newcommand{\Zset}{\ensuremath{\sett{Z}}}

\newcommand{\Zp}{{{\Zset}_{p}}}

\newcommand{\Ztwo}{{{\Zset}_{2}}}

\renewcommand{\vec}[1]{\overrightarrow{#1}}

\newcommand{\Fqt}{\mathcal{F}_{\textrm{QT}}}
\newcommand{\Fdiv}{\mathcal{F}_{\textrm{div}}}

\newcommand{\Fdivgen}{\mathcal{F}_{\textrm{div\_general}}}

\newcommand{\Fmult}{\mathcal{F}_{\textrm{mult}}}
\newcommand{\Fmodconv}{\mathcal{F}_{\textrm{mod}}}
\newcommand{\Fbitdecomp}{\mathcal{F}_{\textrm{BDC}}}
\newcommand{\Fbitcomp}{\mathcal{F}_{\textrm{BC}}}
\newcommand{\Fmsnzbfit}{\mathcal{F}_{\textrm{msnzbfit}}}

\newcommand{\Fextsignabs}{\mathcal{F}_{\textrm{extsignabs}}}
\newcommand{\Fcondassign}{\mathcal{F}_{\textrm{CondAssign}}}
\newcommand{\Finvsqrt}{\mathcal{F}_{\textrm{InvSqrt}}}
\newcommand{\Finvers}{\mathcal{F}_{\textrm{Inv}}}
\newcommand{\Fbdc}{\mathcal{F}_{\textrm{BDC}}}
\newcommand{\Fdivpriv}{\mathcal{F}_{\textrm{divpriv}}}
\newcommand{\Fsqrt}{\mathcal{F}_{\textrm{sqrt}}}
\newcommand{\Fexp}{\mathcal{F}_{\textrm{exp}}}

\newcommand{\Fcompare}{\mathcal{F}_{\textrm{compare}}}
\newcommand{\Fcondassignshare}{\mathcal{F}_{\textrm{CondAssignShare}}}

\newcommand{\smallstart}[1]{
\smallskip
\noindent
\textbf{#1}
}

\newcommand{\zerostartit}[1]{
\noindent
\textit{#1}
}
\newcommand{\smallstartit}[1]{
\smallskip
\noindent
\textit{#1}
}

\usepackage{adjustbox}
\usepackage{threeparttable}
\usepackage{multirow}
\usepackage{wasysym}
\newcommand{\ok}{\CIRCLE} 
\newcommand{\mb}{\LEFTcircle} 
\newcommand{\nn}{\Circle} 
\newcommand\TT{\rule{0pt}{2.6ex}}       
\newcommand\BB{\rule[-1.2ex]{0pt}{0pt}} 

\begin{document}
\title{Adam in Private: Secure and Fast Training of Deep Neural Networks with Adaptive Moment Estimation}
\author{
\IEEEauthorblockN{
Nuttapong Attrapadung\IEEEauthorrefmark{1}, 
Koki Hamada\IEEEauthorrefmark{2}, 
Dai Ikarashi\IEEEauthorrefmark{2}, 
Ryo Kikuchi\IEEEauthorrefmark{2},
Takahiro Matsuda\IEEEauthorrefmark{1}, \\
Ibuki Mishina\IEEEauthorrefmark{2},
Hiraku Morita\IEEEauthorrefmark{3} and 
Jacob C. N. Schuldt\IEEEauthorrefmark{1}
}
\IEEEauthorblockA{
\IEEEauthorrefmark{1}National Institute of Advanced Industrial Science and Technology\\
Email: \{n.attrapadung, t-matsuda, jacob.schuldt\}@aist.go.jp
}
\IEEEauthorblockA{
\IEEEauthorrefmark{2}NTT\\
Email: kikuchi\_ryo@fw.ipsj.or.jp, \{koki.hamada.rb, dai.ikarashi.rd, ibuki.mishina.br\}@hco.ntt.co.jp}
\IEEEauthorblockA{
\IEEEauthorrefmark{3}University of St. Gallen\\
Email: hiraku.morita@unisg.ch}
}


\maketitle

\begin{abstract}
Machine Learning (ML) algorithms, especially deep neural networks (DNN), have proven themselves to be extremely useful tools for data analysis, and are increasingly being deployed in systems operating on sensitive data, such as recommendation systems, banking fraud detection, and healthcare systems. 
This underscores the need for privacy-preserving ML (PPML) systems, and has inspired a line of research into how such systems can be constructed efficiently. 
We contribute to this line of research by proposing a framework 
that allows efficient and secure evaluation of full-fledged state-of-the-art ML algorithms via secure multi-party computation (MPC). 
This is in contrast to most prior works on PPML, which require advanced ML algorithms to be substituted with approximated variants that are ``MPC-friendly'', before MPC techniques are applied to obtain a PPML algorithm. 
A drawback of the latter approach is that it requires careful fine-tuning of the combined ML and MPC algorithms, and might lead to less efficient algorithms or inferior quality ML (such as lower prediction accuracy).
This is an issue for secure training of DNNs in particular, as this involves several arithmetic algorithms that are thought to be ``MPC-unfriendly'', namely, integer division, exponentiation, inversion, and square root extraction.

In this work, we propose secure and efficient protocols for the above seemingly MPC-unfriendly computations (but which are essential to DNN). 
Our protocols are three-party protocols in the honest-majority setting, and we propose both passively secure and actively secure with abort variants.
A notable feature of our protocols is that they simultaneously provide high accuracy and efficiency.
This framework enables us to efficiently and securely compute modern ML algorithms such as Adam (Adaptive moment estimation) and the softmax function ``as is'', without resorting to approximations. As a result, we obtain secure DNN training that outperforms state-of-the-art three-party systems; 
our \textit{full} training is up to $6.7$ times faster than just the \textit{online} phase of the recently proposed FALCON (Wagh et al. at PETS'21) on the standard benchmark network for secure training of DNNs.
To further demonstrate the scalability of our protocols, we perform measurements on real-world DNNs, AlexNet and VGG16, which are complex networks containing millions of parameters.
The performance of our framework for these networks is up to a factor of about $12\sim 14$ faster for AlexNet and $46\sim 48$ faster for VGG16 to achieve an accuracy of $70\%$ and $75\%$, respectively, when compared to FALCON. 

\end{abstract}

\section{Introduction}

Secure multi-party computation (MPC)~\cite{FOCS:Yao82b,STOC:GolMicWig87,STOC:BenGolWig88} enables function evaluation, while keeping the input data secret. An emerging application area of secure computation is privacy-preserving machine learning (ML), such as (secure) deep neural networks. Combining secure computation and deep neural networks, it is possible to gather, store, train, and derive predictions based on data, which is kept confidential. This provides data security and encourages data holders to share their confidential data for machine learning. As a consequence, it becomes possible to use a large amount of data for model training and obtain accurate predictions. 

We first briefly review a typical \emph{training} (or learning) process of a deep neural network in the clear (\ie without secure computation). A deep neural network (DNN) consists of several layers, and certain functions are sequentially computed on the training data layer-by-layer. The so-called \emph{softmax} function is one of the more common functions computed in the last layer. Then, a tentative output from the last layer is computed, and a convergence test is applied to this. 
Based on the result of the test,
the  parameters  are  updated  by  an  optimization  method,  and  the  above  processes  will  be  repeated. A traditional optimization method is \emph{stochastic gradient descent (SGD)}. As SGD tends to incur many repetitions (and hence slow convergence), more efficient approaches have been proposed; adaptive gradient methods such as \emph{adaptive moment estimation (Adam)}~\cite{DBLP:journals/corr/KingmaB14} are popular optimization methods which improve upon SGD and are adopted in many real-world tool-kits, \eg\cite{scikitlearn}. 

A key challenge towards privacy-preserving ML, especially for DNN, is how to securely compute functions that 
are \emph{not} ``MPC-friendly''. MPC-friendly functions refer to functions that are easy to securely compute in MPC, and for which very efficient protocols exist. 
However, unfortunately, functions required in DNN are often \emph{MPC-unfriendly}, especially those used in more modern approaches to training. In particular, Adam~\cite{DBLP:journals/corr/KingmaB14} (and also the softmax function) consist of several MPC-unfriendly functions, namely, integer division, exponentiation, inversion, and square root computations.

To cope with this challenge, up to now, there have been two lines of research. First, many works (to name just a few,~\cite{DBLP:conf/icml/Gilad-BachrachD16,CCS:LJLA17,EPRINT:CGRST17,C:BMMP18,ASIACCS:RWTSSK18,DBLP:conf/dac/RouhaniRK18,DBLP:conf/ccs/ChaudhariCPS19,DBLP:journals/access/KitaiCYNOUTAMH19,DBLP:conf/ndss/PatraS20,PETS:BCPS20,DBLP:journals/iacr/KotiPPS20}) have focused mainly on secure protocols for the \emph{prediction} (or inference) process only, which is much more lightweight compared to the \emph{training}, as gradient optimization methods are not required for prediction. Second, and more recently, there have been a few works in the literature that can handle secure training. These are done mostly by replacing originally MPC-unfriendly functions with different ones that are \emph{MPC-friendly} and approximate the original function on the domain of interest. These approximation approaches either can be done only for elementary optimization methods such as SGD, as in~\cite{DBLP:conf/sp/MohasselZ17,PoPETS:WagGupCha19,CCS:MohRin18,DBLP:conf/ndss/ChaudhariRS20} or require specific ``fine-tuning'' of the interaction between ML and MPC, as in~\cite{CCS:ASKG19}, such that the replaced functions will not degrade the quality of ML architectures significantly (such as lowering prediction accuracy).
In practice, however, this replacement is not easy. 
For example, Keller and Sun \cite{keller2020effectiveness} reported that ASM, which is widely used as a replacement for the softmax function, reduces accuracy in training, sometimes significantly.

Due to the rapid advancements in ML, we believe that a more robust approach to privacy-preserving ML is to achieve efficient protocols for a set of functions that are often used in ML but might typically be thought of as MPC-unfriendly. In this way, the requirement for fine-tuning between ML and MPC would be only minimal, if any at all, and one would be able to plug-and-play new ML advancements into an existing MPC framework to obtain new privacy-preserving ML protocols, without having to worry about the degradation on the ML side.




\subsection{Our Contributions}
We present a framework that allows seamless implementation of secure training for DNNs using modern ML algorithms. Specifically, our contribution is twofold as follows.


\smallstart{New Elementary Three-party Protocols.}
We propose new secure and efficient protocols for a set of elementary functions that are useful for DNN but are normally deemed to be MPC-unfriendly. 
These include secure division, exponentiation, inversion, and square root extraction.
Our protocols are three-party protocols in the honest-majority setting, and we propose both passively secure and actively secure with abort variants. 
A notable feature of our protocols is that they simultaneously provide high accuracy and efficiency. 
A key component to this is our new division protocol, which enables secure fixed-point arithmetic. 
To the best of our knowledge, all previous direct fixed-point arithmetic protocols introduce errors with some probability which must be mitigated, typically resulting in an increased overhead or reduced accuracy. 
In contrast, no such mitigation step is required for our protocols.  
Combined with a range of optimizations suitable for each of the functionalities we consider, we obtain a set of protocols that are both very efficient and ensure high accuracy.
In fact, our efficient implementations of our protocols provide 23-bit accuracy fixed-point arithmetic, which is comparable to single-precision real number operations \emph{in the clear}.
We discuss our construction techniques further in the following section.

\smallstart{New Applications to ML.} We apply our new elementary MPC protocols to ``seamlessly'' instantiate secure computations for softmax and Adam. That is, due to our elementary MPC protocols, we can securely and efficiently compute softmax and Adam ``as is'', in particular, without approximation using (MPC-friendly) functions. Consequently, due to the fast convergence of Adam, we obtain fast and secure training (and prediction) protocols for DNN. 
Using the DNN architecture and MNIST dataset typically used as a benchmark, our protocol achieved 95.64\% accuracy within 117 seconds, improving upon the state-of-the-art such as ABY3~\cite{CCS:MohRin18} (94\% accuracy within 2700 seconds reported in \cite{CCS:MohRin18}) and FALCON~\cite{Falcon} (780 seconds for the online phase only)\footnote{The measurements for our protocol and FALCON were done in the environment described in Section~\ref{sec:experiment}, which is roughly comparable to the one in \cite{CCS:MohRin18}.}.  
Moreover, our training converges much faster, namely, in one epoch, as opposed to 15 epochs for ABY3 and FALCON.
Furthermore, our protocol achieves the same accuracy as training over \emph{plaintext} data, using MLPClassifier in the scikit-learn tool-kit~\cite{scikit-learn} while being less than six times slower. 
We further perform measurement on real-world DNNs from the ML literature, AlexNet~\cite{AlexNet} and VGG16~\cite{VGG16}, which contain millions of parameters.
Comparing the \emph{total training time} (i.e. time to reach a certain accuracy),
the total running time of our framework outperforms the online phase of FALCON with a factor of about $12\sim14$ for AlexNet and $46\sim48$ for VGG16 in the LAN setting. 
A detailed performance evaluation and comparison considering different security and network settings, different datasets, and large DNNs, is given in Section~\ref{sec:experiment}.  


\subsection{Our Techniques}

\smallstart{New Techniques for Secure Truncation.} 
We first briefly describe the idea behind a common building block for all our protocols: division (which also implies truncation). Let $p$ be the size of the underlying ring/field, $x$ be the secret and $d$ is the divisor (so the desired output is $\frac{x}{d}$). Known efficient truncation protocols, \eg\cite{CCS:MohRin18,DBLP:conf/sp/MohasselZ17},  reconstruct a masked secret $x+r$ for a random $r$, divide this by $d$ in the clear, and subtract $\frac{r}{d}$. However, in this approach, a large error, $-\frac{p}{d}$, sneaks into the output when $x+r > p$ because the reconstructed value becomes $x+r-p$. To avoid this, the message space has to be much smaller than $p$, which leads to reduced accuracy for a given value of $p$.
Instead, we employ a different approach. Let $x_1$ and $x_2$ be additive shares of $x$ such that $x_1 + x_2 = x+ qp$ for $q\in \{0,1\}$. Our approach is to securely compute $q$ and eliminate $qp$ (without exposing $q$ to any parties), which makes the (local) division of sub-shares be the desired output. In this way, we can embed a large value into a single share, which, in turn, enables accurate computation of functions such as exponentiation.

\smallstart{New Techniques for Elementary Protocols.} For securely computing exponentiation, inversion, division with private divisor, square root, and inversion of square root, we utilize Taylor or Newton series expansions. 
A key challenge here is to ensure fast convergence that, in general, is only guaranteed for a narrow range of input values. 
We resolve this by constructing protocols that use a combination of private input pre-processing and partial evaluation of the pre-processed input.   
We devise \emph{private scaling} techniques, which allow inputs to be scaled to fit an optimal input range, and furthermore allow the protocol to make the most out of the available bit range in the internal computations.
We also utilize what we call \emph{hybrid table-lookup/series-expansion} techniques, which separate inputs into two parts and apply table-lookup and series-expansion to the respective parts. 
The details of how these techniques are used in our protocols differ depending on the functionality of the protocols. 
We provide detailed descriptions in Section~\ref{section:elementary}.

\subsection{Related Work}
\begin{table*}[t]
\centering
\caption{ Comparison among various privacy-preserving ML systems. \label{tab:allmlcomparison} } 
\begin{adjustbox}{max width=\textwidth}
\begin{threeparttable}
\begin{tabular}{cccccccccccccccccccccccccccccccc} 
\toprule
&& 
\rotatebox[origin=l]{80}{Prediction} & 
\rotatebox[origin=l]{80}{Training} &
\,
&
\rotatebox[origin=l]{80}{Basic} & 
\rotatebox[origin=l]{80}{Batch-Norm} & 
\rotatebox[origin=l]{80}{Advance (ex. Adam)} &
\,
&
\rotatebox[origin=l]{80}{Semi-honest} & 
\rotatebox[origin=l]{80}{Malicious} &
\,
&
\rotatebox[origin=l]{80}{HE} & 
\rotatebox[origin=l]{80}{GC} &
\rotatebox[origin=l]{80}{SS} &
\,
&
\rotatebox[origin=l]{80}{LAN} & 
\rotatebox[origin=l]{80}{WAN} &
\,
&
\rotatebox[origin=l]{80}{Small (ex. MNIST)} & 
\rotatebox[origin=l]{80}{Large (ex. CIFAR-10)} &
\,
&
\rotatebox[origin=l]{80}{Simple (ex. 3DNN)} & 
\rotatebox[origin=l]{80}{Complex (ex. VGG-16)} &
\,
\\
\cline{3-4}
\cline{6-8}
\cline{10-11}
\cline{13-15}
\cline{17-18}
\cline{20-21}
\cline{23-24}
\TT\BB & System &
\multicolumn{2}{c}{
\begin{tabular}{c} \TT Secure \\ Capability \BB \end{tabular}
}
&&
\multicolumn{3}{c}{
\begin{tabular}{c} \TT Supported ML\\ Algorithms \BB \end{tabular}
}
&&
\multicolumn{2}{c}{
\begin{tabular}{c} \TT Threat \\ Model \BB \end{tabular}
}
&&
\multicolumn{3}{c}{
\begin{tabular}{c} \TT Based \\ Techniques \BB \end{tabular}
}
&&
\multicolumn{2}{c}{
\begin{tabular}{c} \TT LAN/ \\ WAN \BB \end{tabular}
}
&&
\multicolumn{2}{c}{
\begin{tabular}{c} \TT Evaluation \\ Dataset \BB \end{tabular}
}
&&
\multicolumn{2}{c}{
\begin{tabular}{c} \TT Network \\ Architectures \BB \end{tabular}
}
\\
\cline{3-15}
\cline{17-24}
\TT\BB
&& \multicolumn{13}{c}{Theoretical metric} && \multicolumn{8}{c}{Evaluation metric}
\\
\toprule
\multirow{7}{*}{2PC}
& MiniONN~\cite{CCS:LJLA17}
& \ok & \nn && \ok & \nn & \nn && \ok & \nn && \ok & \ok & \ok
&& \ok & \nn && \ok & \ok  && \ok & \nn 
\\
& Chameleon~\cite{ASIACCS:RWTSSK18}
& \ok & \nn && \ok & \nn & \nn && \ok & \nn && \nn & \ok & \ok
&& \ok & \ok && \ok & \ok  && \ok & \mb
\\
& EzPC~\cite{EPRINT:CGRST17}
& \ok & \nn && \ok & \nn & \nn && \ok & \nn && \nn & \ok & \ok
&& \ok & \ok && \ok & \ok  && \ok & \nn 
\\
& Gazelle~\cite{USENIX:JuvVaiCha18}
& \ok & \nn && \ok & \nn & \nn && \ok & \nn && \ok & \ok & \ok
&& \ok & \nn && \ok & \ok  && \ok & \nn 
\\
& SecureML~\cite{DBLP:conf/sp/MohasselZ17}
& \ok & \ok && \ok & \nn & \nn && \ok & \nn && \ok & \ok & \ok
&& \ok & \mb && \ok & \nn  && \ok & \nn 
\\
& XONN~\cite{USENIX:RSCLLK19}
& \ok & \nn && \ok & \mb & \nn && \ok & \ok && \nn & \ok & \ok
&& \ok & \nn && \ok & \ok  && \ok & \mb 
\\
& Quotient~\cite{CCS:ASKG19}
& \ok & \ok && \ok & \ok & \ok && \ok & \nn && \nn & \ok & \ok
&& \ok & \ok && \ok & \nn  && \ok & \mb 
\\
& Delphi~\cite{USENIX:MLSZP20}
& \ok & \nn && \ok & \nn & \nn && \ok & \nn && \ok & \ok & \ok
&& \ok & \nn && \nn & \ok  && \ok & \mb
\\
& FHE-based SGD~\cite{DBLP:conf/cvpr/NandakumarRPH19}
& \ok & \ok && \ok & \ok & \nn && \ok & \nn && \ok & \nn & \nn
&& \ok & \nn && \ok & \nn  && \ok & \nn
\\
& Glyph~\cite{DBLP:conf/nips/LouFF020}
& \ok & \ok && \ok & \ok & \nn && \ok & \nn && \ok & \nn & \nn
&& \ok & \nn && \ok & \ok  && \ok & \nn
\\
\midrule
\multirow{8}{*}{3PC}
& ABY3~\cite{CCS:MohRin18}
& \ok & \ok && \ok & \nn & \nn && \ok & \ok && \nn & \ok & \ok
&& \ok & \mb && \ok & \nn  && \ok & \nn 
\\
& SecureNN~\cite{PoPETS:WagGupCha19}
& \ok & \ok && \ok & \mb & \nn && \ok & \mb && \nn & \nn & \ok
&& \ok & \ok && \ok & \nn  && \ok & \nn 
\\
& CryptFlow~\cite{SP:KRCGRS20}
& \ok & \nn && \ok & \nn & \nn && \ok & \ok && \nn & \nn & \ok
&& \ok & \nn && \ok & \ok  && \ok & \mb 
\\
& QuantizedNN~\cite{EPRINT:DalEscKel20}
& \ok & \nn && \ok & \mb & \nn && \ok & \ok && \ok & \nn & \ok
&& \ok & \ok && \nn & \ok  && \nn & \mb
\\
& ASTRA~\cite{DBLP:conf/ccs/ChaudhariCPS19}
& \ok & \nn && \ok & \nn & \nn && \ok & \ok && \nn & \ok & \ok
&& \ok & \ok && \ok & \nn  && \ok & \nn 
\\
& BLAZE~\cite{DBLP:conf/ndss/PatraS20}
& \ok & \mb && \ok & \nn & \nn && \ok & \ok && \nn & \ok & \ok
&& \ok & \ok && \ok & \nn  && \ok & \nn 
\\
& Falcon~\cite{Falcon}
& \ok & \ok && \ok & \ok & \nn && \ok & \ok && \nn & \nn & \ok
&& \ok & \ok && \ok & \ok  && \ok & \ok 
\\
& \textbf{This work}
& \ok & \ok && \ok & \ok & \ok && \ok & \ok && \nn & \nn & \ok
&& \ok & \ok && \ok & \ok  && \ok & \ok 
\\
\midrule
\multirow{2}{*}{4PC}
& FLASH~\cite{PETS:BCPS20}
& \ok & \nn && \ok & \nn & \nn && \ok & \ok && \nn & \nn & \ok
&& \ok & \ok && \ok & \mb  && \ok & \nn 
\\
& Trident~\cite{NDSS:RachuriS20}
& \ok & \ok && \ok & \nn & \nn && \ok & \ok && \nn & \ok & \ok
&& \ok & \ok && \ok & \nn  && \ok & \nn 
\\
\bottomrule
\end{tabular}
{\small
``Basic'' for Supported ML Algorithms refers to more basic ones such as linear operations, convolution, ReLU, Maxpool, and/or SGD optimizer.
 ``Advance'' refers to advance optimizers, namely, ADAM (considered in this work) and AMSGrad (in Quotient). 
HE, GC, SS refer to homomorphic encryption, garbled circuit, and secret sharing, respectively.
``Small'' for Evaluation Dataset refers to MNIST, except for BLAZE, which uses Parkinson disease dataset and for Quotient, which uses also MotionSense, Thyroid, and more, besides MNIST (their dimensions are similar to MNIST).
``Large'' refers to larger datasets such as the well-known CIFAR-10 in particular (in all the systems that tick except QuantizedNN), or TinyImageNet (in CryptFlow and QuantizedNN, and partially in Falcon). 
``Simple'' for Network Architectures refers to simple neural networks such as the basic 3-layer DNN (3DNN) from SecureML in particular, or other slightly different small networks from~\cite{CCS:LJLA17,PoPETS:WagGupCha19}.
``Complex'' refers to more complex networks such as the well-known AlexNet and VGG-16 in particular (both are considered in Falcon and this work, while XONN uses VGG-16 among other networks).
$\ok$ indicates that such a system support a feature,
$\nn$ indicates that such a system does not so support so,
$\mb$ refers to fair comparison being difficult due to various reasons.
Secure training in BLAZE only considers the case for less advance ML algorithms \eg linear/logistic regression, but notably not neural networks. (Their secure prediction, on the other hand, includes neural networks).
XONN, QuantizedNN support simplified or different versions of batch normalization, while SecureNN supports divisions but not batch norm.
SecureML only estimate their WAN evaluation, while
ABY3 does not present WAN results for neural networks. 
SecureNN achieves malicious privacy (but not including correctness), as defined in~\cite{CCS:AFLNO16}.
FLASH uses a smaller data set than CIFAR-10.
Networks with moderate sizes are experimented: \eg ResNet-20 (in Quotient), ResNet-32 (in Delphi), ResNet-50, DenseNet-121 (in CryptFlow), MobileNet (in QuantizedNN).
Chameleon uses a slightly weaker version of AlexNet. 
%
}
\end{threeparttable}
\end{adjustbox}
\end{table*}

Various ML algorithms have been considered in connection with privacy preserving ML, include decision trees, linear regression, logistic regression, support-vector-machine classifications, and deep neural networks (DNN). Among these, deep neural networks are the most flexible and have yielded the most impressive results in the ML literature. However, at the same time, secure protocols for DNN are the most difficult to obtain, especially for the training process.
We show a table for comprehensive comparison among PPML systems supporting DNN in Table~\ref{tab:allmlcomparison}.

\smallstart{Secure DNN Training.}
Our work focuses on \emph{secure training for deep neural networks} (secure inference can be obtained as a special case). 
There have been several works on secure DNN training such as SecureML~\cite{DBLP:conf/sp/MohasselZ17}, SecureNN~\cite{PoPETS:WagGupCha19},  ABY3~\cite{CCS:MohRin18}, Quotient~\cite{CCS:ASKG19}, FHE-based SGD~\cite{DBLP:conf/cvpr/NandakumarRPH19}, Glyph~\cite{DBLP:conf/nips/LouFF020}, Trident~\cite{DBLP:conf/ndss/ChaudhariRS20}, and FALCON~\cite{Falcon}.
All of these achieve efficiency by simplifying the underlying DNN training algorithms (e.g. replacing functionalities with less-accurate easier-to-compute alternatives), and optimizing the computation of these. As a consequence of this approach, they are restricted to simple SGD optimization, with the exception of Quotient which implements an approximation to AMSGrad. We emphasize that we take a fundamentally different approach by constructing protocols that allow unmodified advanced training to be done efficiently.
In the following, we highlight properties of the above related works.

\smallstart{Setting/Security.} SecureML, Quotient, FHE-based SGD, and Glyph are two-party protocols, 
SecureNN, ABY3, and FALCON are three-party protocols, while Trident is a four-party protocol in a somewhat unusual asymmetric offline-online setting. SecureML, Quotient, FHE-based SGD, Glyph, and SecureNN considered \emph{semi-honest (passive)} security tolerating one corrupted party, while SecureNN can be extended to achieve so-called \emph{privacy against malicious adversaries} (formalized by~\cite{CCS:AFLNO16}). 
ABY3 improved security upon these by considering \emph{malicious (active) security with abort} tolerating one corrupted party. Trident improved security in term of \emph{fairness} (again, tolerating one corrupted party); this comes with the cost of reducing the tolerated corruption fraction from 33\% to 25\%.
It should be noted that, unlike the other schemes, FALCON sacrifice perfect security to compute batch-normalization more efficiently (see Section \ref{sec:secureprotNN}).

\smallstart{Efficiency.} 
For secure training over a basic 3-layer DNN on the MNIST dataset, ABY3 outperforms both SecureML/SecureNN and was state-of-the-art before Trident and FALCON.
FHE-based SGD and Glyph use fully homomorphic encryption, which makes non-interactive training possible. 
Glyph is the most efficient of the two, but is still far less efficient than ABY3 in terms of execution time.
Trident improves the \emph{online} phase of ABY3 but with the cost of adding a fourth party who only participates in offline phase. 
Most recently, FALCON also improves upon the \emph{online} phase of ABY3.
As highlighted above, our framework improves upon FALCON.

\smallstart{Additional Related Works.} 
 Note that when considering only secure \emph{inference} (but not secure training) for DNN, BLAZE~\cite{DBLP:conf/ndss/PatraS20} achieved stronger security (than that of ABY3) of \emph{fairness}.

For the less flexible ML algorithms, namely, linear/logistic regression (which we do not focus on in this work), there have been recent progresses in secure training. Up to 2018, ABY3 was the state-of-the-art in terms of performance and security, achieving the same security as their DNN counterpart. Recently, for linear/logistic regression, BLAZE~\cite{DBLP:conf/ndss/PatraS20} improved ABY3 by 50-2600 times in performance and also achieved \emph{active security with fairness}. More recently,  SWIFT~\cite{DBLP:journals/iacr/KotiPPS20} achieved the strongest notion, namely, \emph{active security with guaranteed output delivery}. 

\section{Preliminaries and Settings} \label{sec:prelims}

\zerostartit{Notations for Division.} For $a,b\in \Zset$, we denote by $\dfrac{a}{b} \in \mathbb{R}$ real-valued division, 
and by $a/b \in \mathbb{Z}$ integer division that discards the remainder. In other words, $a/b:= \lfloor \dfrac{a}{b} \rfloor$.

\subsection{Data Representation} \label{sec:datarep}
The data representation is an important aspect of efficient and accurate computation. 
The algorithms considered in this paper make use of the following data types:
\begin{itemize} \setlength{\leftskip}{-2mm}
    \item Binary values $\Ztwo$.
    \item $\ell$-bit unsigned and signed integers, $\Zset_{\langle \ell \rangle}^+ = \{a \in \Zset \mid 0 \leq a \leq 2^{\ell} -1 \}$ and $\Zset_{\langle \ell \rangle} = \{a \in \Zset \mid -2^{\ell-1} \leq a \leq 2^{\ell-1} -1 \}$, respectively, where the range of values for signed integers reflect that a single bit is used to indicate the sign.
    \item $\ell$-bit fixed-point unsigned and signed rational numbers $\mathbb{Q}^+_{\langle \ell, u \rangle} = \{ b \in \mathbb{Q} \mid b = \frac{a}{2^u}, a \in \Zset_{\langle \ell \rangle}^+\}$ and $\mathbb{Q}_{\langle \ell, u \rangle} = \{ b \in \mathbb{Q} \mid b = \frac{a}{2^u}, a \in \Zset_{\langle \ell \rangle}\}$, respectively.
\end{itemize}
%
We represent these data types as follows.
Binary values are represented as is, i.e.\ as elements of the field $\Fset = \Ztwo$.
Signed and unsigned integers are represented as elements of the field $\Fset = \Zset_p$ for a Mersenne prime $p > 2^k$.
This implies that 
signed integers are represented using ones' compliment 
(i.e. a negative value $-a \in \Zset_{\langle \ell \rangle}$ is represented as $p -a$, and the most significant bit, which indicates the sign, will be $1$).
We will likewise represent fixed-point values as elements of $\Fset_p$, and in order to do so, these are scaled to become integers.
Specifically, we will use a set of (unsigned) $\ell$-bit integers $0\leq a \leq 2^{\ell} - 1$, which we denote $\widehat{\mathbb{Q}}^+_{\langle \ell, \alpha \rangle}$, to represent the values $\{\frac{a}{2^\alpha} | 0\leq a \leq 2^{\ell} - 1 \}$, and will refer to $\alpha$ as the \textit{offset} for these. 
For a fixed-point value $a$, we will use the notation $a_{\pre{\alpha}}$ to denote the integer representation with offset $\alpha$ i.e. $a_{\pre{\alpha}} = a \cdot 2^\alpha$. 
The integers in $\widehat{\mathbb{Q}}^+_{\langle \ell, \ell \rangle}$ are represented as elements of $\Fset_p$, and we denote the signed extension by $\widehat{\mathbb{Q}}_{\langle \ell, \ell \rangle}$.


Note that the representation of fixed-point values requires the scaling factor to be taken into account for multiplication (and division).
Specifically, for values $a_{\pre{\alpha}}$ and $b_{\pre{\alpha}}$, the correct representation of the product of $a$ and $b$ is $a_{\pre{\alpha}} \cdot b_{\pre{\alpha}} / 2^\alpha = (a \cdot b)_{\pre{\alpha}}$.
For simplicity, we use $\times_\alpha$ to denote this operation, i.e. (ordinary) multiplication followed by division by $2^\alpha$.

Finally note that since $p = 2^\lambda -1$ is a Mersenne prime, modular arithmetic in $\Fset_p$ can be done swiftly via bit-shifting and addition (e.g. see~\cite{BKLM10}).
Specifically, if $a = a_0 2^\lambda + a_1$, then $a \mod p = a_0 2^\lambda + a_1 \mod p = a_0 + a_1 \mod p$ holds since $2^\lambda -1 = 0 \mod p$.
This will allow efficient operations on shared integers or fixed-point values represented as above.

\subsection{Multi-Party Computation Setting}
We consider secret-sharing (SS)-based three-party computation secure against a single static corruption:
There are three parties $P_1, P_2, P_3$, a secret is shared among these parties via SS,
any two parties can reconstruct the secret from their shares,
and an adversary corrupts up to a single party at the beginning of the protocol. For notational convenience, we treat the party index $i \in \Zset$ as to refer to the $i'$-th party where $i' \equiv i \pmod 3$ and $i'\in\{1,2,3\}$. For example, $P_0=P_3$ and $P_4=P_1$.
%

We consider the client/server model.
This model is used to outsource secure computation,
where any number of clients send shares of their inputs to the servers.
Hence, both the input and output of the servers are in a secret-shared form,
and our protocols are thus share-input and share-output protocols.
\rnote{added workflow} More precisely, during secure training, the three parties have shares of training data as input, 
and then interact with each other to obtain \emph{shares} of a trained model.
This setting is composable, i.e., there is a degree of freedom in how the input and output come in and how they are used. For example, a client other than the parties may provide input, or the output of another secure protocol may be used as an input. The resulting model can be made public, or the prediction can be made while maintaining the model secret.

Regarding the adversarial behavior, 
we consider both passive (semi-honest) and active (malicious) adversaries with abort.
In passive security, 
corrupted parties follow the protocol but might try to obtain private information 
from the  transcripts of messages that they receive.
Formally, we say that a protocol is passively secure 
if there is a simulator that simulates the view of the corrupted parties
from the inputs and outputs of the protocol~\cite{Gol01}.
In active security with abort,
corrupted parties are allowed to behave arbitrarily in attempt to break the protocol.
The probability an active adversary successfully cheats 
is parametarized by the statistical security parameter $\kappa$,
meaning that probability is bounded by $2^{-\kappa}$.%
We prove the security of our protocols in a hybrid model, 
where parties run the real protocols, 
but also have access to a trusted party computing specified subfunctionalities for them. 
For a subfunctionality denoted $g$, we say that the protocol runs in the $g$-hybrid model.

\subsection{Secret Sharing Schemes and Their Protocols}\label{sect:ss}
In this paper, we use three \emph{replicated secret sharing schemes}~\cite{GLOBECOM:ItoSaiNis87,CDI05}. We consider the 2-out-of-3 threshold access structure for the first two schemes. For the third scheme, the minimal access structure is simply $\{\{1,2\}\}$, meaning only $P_1$ and $P_2$ can together reconstruct the secret.
We~denote~them~as:
\begin{align*}
	\begin{aligned}
	\pash{\cdot} &\text{-sharing} &&\text{: the 2-out-of-3 replicated sharing in~$\Zset_p$}, \\
	\bash{\cdot} &\text{-sharing} &&\text{: the 2-out-of-3 replicated sharing in~$\Zset_2$}, \\
	\tash{\cdot} &\text{-sharing} &&\text{: the simple additive sharing in~$\Zset_p$}.
	\end{aligned}
\end{align*}

%

\noindent\underline{$\pash{\cdot}$-sharing}. This scheme is specified by:
\begin{itemize} \setlength{\leftskip}{-2mm}
	\item Share: 
    	To share $a \in \mathbb{Z}_p$, pick random $a_1, a_2, a_3 \in \mathbb{Z}_p$ such that $a = a_1+a_2+a_3$, then set $\pash{a}_i = (a_i,a_{i+1})$ as the $P_i$'s share for $i = 1,2,3$. Denote $\pash{a}=(\pash{a}_1, \pash{a}_2, \pash{a}_3)$.
    \item Reconstruct:
        Given a pair of shares of $a$,
        this protocol with passive adversary guarantees that all the parties eventually obtain $a$.
        With an active adversary, this functionality proceeds the same unless $\pash{a}$ is not consistent, where all the honest parties will abort at the end of the execution.

     \item Local operations:
		Given shares $\pash{a}$ and $\pash{b}$ and a scalar 
		$\alpha \in \mathbb{Z}_p$,
        the parties can generate shares of $\pash{a+b}$, $\pash{\alpha a}$, and $\pash{\alpha + a}$
        using only local operations.
        The notations $\pash{a}+\pash{b}$, $\alpha \pash{a}$, and $\alpha + \pash{a}$ denote these local operations, respectively.
      \item Multiplications: 
Given shares $\pash{a}$ 
and $\pash{b}$, 
the parties can generate $\pash{ab}$ 
by a multiplication protocol~\cite{DN07,GRR98,DBLP:conf/ccs/ChidaHIKP18}.
We denote this functionality as $\Fmult$.
\end{itemize}
\noindent\underline{$\bash{\cdot}$-sharing}. This is exactly as $\pash{\cdot}$-sharing but with $p=2$.

Combining local operations with multiplication protocols,
the parties can compute any arithmetic/boolean circuit over shared data, \eg 
the parties obtain $\bash{a \wedge b}$ by multiplying $\bash{a}$ and $\bash{b}$ via $\Fmult$.

\noindent\underline{$\tash{\cdot}$-sharing}. To share $a \in \mathbb{Z}_p$, pick random $\tash{a}_1, \tash{a}_2\in \mathbb{Z}_p$ such that $a = \tash{a}_1 + \tash{a}_2$. Set $\tash{a}_3$ as the empty string. $\tash{a}_i$ is the $P_i$'s share.
Denote $\tash{a}=(\tash{a}_1, \tash{a}_2, \tash{a}_3)$.

\smallstartit{Share Conversions.} Our protocols will utilize conversions among sharing types. Due to limited space, we defer the details to Appendix~\ref{sec:shareconversion}, and provide a summary in Table~\ref{tab:conversion} below. Here, for $a \in \Zset_p$ we let $(a_\ell,\ldots,a_1)$ be the bit representation of $a$; that is, $a = \sum_{i=1}^{\ell}2^{i-1}a_i$. Round is of a passively secure protocol.

\begin{table}[htb]\footnotesize
\caption{Share Conversions \label{tab:conversion}}
\vspace{-3mm}
    \centering
    \begin{tabular}{@{}l@{\ \,}l@{\ }l@{\ }c}
\toprule
Conversion & Functionality name & Protocol & Round \\
\hline
$\pash{a} \to \tash{a}$ 
&
$\ptotconvert$
&
local operations
&
0
\\
$\tash{a} \to \pash{a}$
&
$\ttopconvert$
&
one $\pash{\cdot}$-sharing
&
1
\\
$\pash{a} \to (\bash{a_\ell},\ldots,\bash{a_1})$
&
$\Fbitdecomp$ -- Bit decomposition
&
\cite{ACISP:KIMHC18}
&
$\ell+1$
\\
$(\bash{a_\ell},\ldots,\bash{a_1}) \to \pash{a}$
&
$\Fbitcomp$ -- Bit composition
&
\cite{CCS:ABFKLO18} (modified)
&
$\ell+1$
\\
$\bash{a} \to \pash{a}$
&
$\Fmodconv$ -- Modulus conversion
&
\cite{ACISP:KIMHC18}
&
1
\\
\bottomrule
    \end{tabular}
\end{table}

\smallstartit{Conditional Assignment.}
We define a functionality of conditional assignment $\pash{z} \leftarrow \Fcondassign(a,b,\bash{c})$
via setting $z:=a$ if $c=0$ and $z:=b$ if $c=1$. 
A protocol for this simply converts $\pash{c} := \Fmodconv(\bash{c})$, and computes $\pash{z} := a \cdot (1 - \pash{c}) + \pash{c} \cdot b$.

\subsection{Quotient Transfer Protocol}
Consider the reconstruction of shared secret, $\pash{a}$ or $\tash{a}$, over $\NN$ as opposed to $\Zset_p$ for which $\pash{\cdot}$ and $\tash{\cdot}$ sharings are defined.
The resulting value would be of the form $a + qp$. 
We will refer to $q$ as the \emph{quotient}.
The ability to compute $q$ (in shared form), which we refer to as quotient transfer, will play an important role in our division protocol.

Kikuchi\etal~\cite{ACISP:KIMHC18} proposed efficient three-party quotient transfer protocols for passive and active security. 
In this paper, we use their protocols specialized to the following setting;
firstly, we use a Mersenne prime $p$ for the field $\Zset_p$ underlying the sharings, and we will use $\tash{\cdot}$-sharing for passive security and $\pash{\cdot}$-sharing for active security. 
In this setting, $a$ is required to  be a multiple of $2$ and $4$ in the presence of passive and active adversaries, respectively.
We address this by simply multiplying $\tash{a}$ and $\pash{a}$ by $2$ and $4$, respectively, before conducting the quotient transfer protocols.
It means that a secret $a$ should satisfy $2a < p$ and $4a < p$ for passive and active security, respectively.
%
These protocols have been implicitly used as building blocks for other protocols in \cite{ACISP:KIMHC18},
but were not explicitly defined. 
We thus describe them in Appendix~\ref{sect:QTprotocol},
as well as a quotient transfer protocol for \mbox{$\bash{\cdot}$-sharings},
which is a popular way to obtain the carry for binary addition.

The quotient transfer functionality, $\Fqt$, is defined in Functionality~\ref{func:qt}.
For brevity,
$\Fqt$ is defined for $\tash{a}$, but we additionally use this functionality for $\pash{a}$ and $\bash{a}$. 
Note that for $\tash{a}$, 
$q \in \{0,1\}$ where $\tash{a}_1 + \tash{a}_2 = a + qp$; 
for $\pash{a}$, $q \in \{0,1,2\}$ where sub-shares $a_1 + a_2 + a_3 = a + qp$; and for $\bash{a}$,
and $q \in \{0,1\}$ where sub-shares $a_1 + a_2 + a_3 = a + 2q$.


\begin{figure}[htbp]
\centering
\framebox[\width][c]{
    \small\,
    \hbox{
    \begin{varwidth}[c]{0.43\textwidth}
    \begin{myfunctionality}
    [$\Fqt$ -- Quotient transfer]
    \label{func:qt}
    \end{myfunctionality}
    {Upon receiving $\tash{a}$, $\Fqt$ sets $q$ such that $\tash{a}_1 + \tash{a}_2 = a + qp$ in $\NN$,
    generates shares $\tash{q}$, and sends $\tash{q}_i$ to $P_i$.}
    \end{varwidth}
    \,
    }
}
\end{figure}

\section{Secure Real Number Operations}
\label{section:securerealnumber}
In this section, we present the division protocols that will allow us to do fixed-point arithmetic efficiently and securely.
The key to achieve efficient fixed-point computations is the ability to
perform \emph{truncation} (or equivalently, integer division by $2^k$, also called right-shift), 
as this allows multiplication of scaled integer representations of fixed-point values, as introduced in Section~\ref{sec:datarep}.  

Our new division protocol is efficient and accurate.
The popular division (truncation) protocol approach used in the context of machine learning requires a heavy offline phase, 
although it is efficient in the online phase. 
In addition, as we will see later, there is a possibility of introducing errors that is much larger than rounding errors. 
Hence, we present a protocol that is efficient in its overall cost, i.e., the total cost is comparable only to the current \emph{online} cost, 
while eliminating the possibility of introducing large errors.

\subsection{Current Secure Division Protocol}\label{sec:currentdiverror}
In this section, we analyze the approach taken to division in current multi-party computation protocols \cite{CCS:MohRin18,PoPETS:WagGupCha19,NDSS:RachuriS20}
and show why the large error can be introduced in the output.
For simplicity, we consider \emph{unsigned} integers shared over $\Zset_p$ for a general $p$ and a general divisor $d$,
but similar observations holds for the signed integers and specific $p$, such as $2^{64}$.

Let $(a_1,a_2,a_3)$ be the sub-shares of $a$ in the replicated secret sharing scheme and $a = \alpha_a d + r_a$ for $0 \le r_a < d$,
and let $(b_1,b_2,b_3)$ be the sub-shares of the output of a division protocol.
Here, the intention is that $b_1+b_2+b_3$ is a value close to $\frac{a}{d}$,
such as $\alpha_a$, or perhaps  $\alpha_a \pm 1$.

The typical protocol proceeds as follows.
The parties first prepare a shared correlated randomness $(\pash{s'},\pash{s})$, 
where $s' \gets \Zp$ and $s := \intdiv{s'}{d}$. (Note that $d$ is public and known a priori.)
The parties then compute $\pash{a+s'}$, reconstruct $(a+s')$, and set $\pash{b} = \pash{s} + \intdiv{(a+s')}{d}$.

This protocol seems to work well, but the output can in fact be far from the intended $\frac{a}{d}$.
To see this, let 
$s' = s d + r_{s'}$ and $p = \alpha_p d + r_p$.
Considering the reconstructed value $\pash{a+s'}$ over $\NN$, we see that the parties obtain $a + s' - qp$, 
where $q \in \{0,1\}$. 
Hence, the computed shared secret corresponds to 
\begin{align}
 \lefteqn{s + \intdiv{(a + s' - qp)}{d}} \nonumber \\
&= s + \intdiv{\left((\alpha_{a}d+r_a) - (sd+r_{s'}) -q (\alpha_pd+r_p)\right)}{d}  \nonumber \\
&= \alpha_{a} - q \alpha_p + \intdiv{(r_a - r_{s'} + qr_p)}{d} \label{eq:aby2out}.
\end{align}

Roughly speaking, the third term, $\intdiv{(r_a - r_{s^\prime} - qr_p)}{d}$, 
is a small constant since $r_a$, $r_{s^\prime}$, and $r_p$ are less than $d$.
Hence, this term can be considered to be a small rounding error.
On the other hand, the second term, $q \alpha_p$, can be large if $q = 1$. 
For example, if we set $p=2^{64}$ and $d=2^{12}$, then $\alpha_p = 2^{52}$, and the reconstructed result will differ from $\frac{a}{d}$ by $2^{52}$.
To address this, we have to make the probability of $q=1$ negligible.
If a protocol conducts several divisions with $p=2^k$ and $\ell$-bit inputs,
the probability that $q=1$ occurs at least once during the $\tau$ divisions
is $1 - \left(1-2^{-k+\ell}\right)^\tau$.
It is likely to happen if we conduct the division protocol many times.
For example, this probability is over $90\%$ on 15 epochs MNIST training with the batch-size 128, $k=64$, and $\ell=36$.
A single event of $q=1$ does not necessarily lead to a catastrophic error in the intended function; 
however, many occurrences of $q=1$ can make the result differ significantly from the intended~value. 

To keep the probability of an error occurring below a given threshold,
the input space (i.e. the parameter $\ell$) can be adjusted to be sufficiently small such that the required number of divisions can be accommodated.
However, setting $\ell$ depending on the required number of divisions is often problematic, since this number can be difficult to estimate a priori in an exploratory analysis such as machine learning.
Hence, a common approach is to set $\ell$ small enough to ensure $q=0$ with overwhelming probability. 
This, however, leads to a larger reduction of the input space, which can negatively impact the computation being done, due to lower supported accuracy.

\subsection{Our Protocol for Division by Public Value}

\subsubsection{Intuition}\label{sec:intuition}
We first give the intuition behind our protocols.
In our protocol for input $\pash{a}$, we locally convert $\pash{a}$ into $\tash{a}$ before division.
Hence, in the following, we assume the input is $\tash{a}$ and a public divisor $d$. 

First, let us analyze what happens when we simply divide each share by $d$.
Let $\tash{a}_1 + \tash{a}_2 = qp+a$ in $\NN$, where $q \in \{0,1\}$.
Here, suppose that $\tash{a}_j = \alpha_j d + r_j$, $a = \alpha_a d + r_a$, and $p = \alpha_p d + r_p$ for $j=1,2$ 
If each party divide its share $\tash{a}_j$ by $d$, 
the new share is $\alpha_j$, \ie $\intdiv{\tash{a}_j}{d} = \alpha_j$.
Then, the reconstruction of $(\alpha_1,\alpha_2)$ will be
\begin{align}\label{eq:simpledivision}\small
    \alpha_1+\alpha_2 = \alpha_a + q\alpha_p + \frac{r_a+qr_p - (r_1+r_2)}{d},
\end{align}
which contains extra terms, $q\alpha_p$ and $\frac{(r_a+qr_p - (r_1+r_2))}{d}$.

The insight behind our protocols is that the $q\alpha_p$ term can be eliminated, which we essentially achieve via the quotient transfer protocol~\cite{ACISP:KIMHC18}, that allow us to obtain $\tash{q}$ efficiently.
This protocol suits our setting since it requires a prime $p$, and prefers a Mersenne prime.
The quotient transfer protocol furthermore requires $a$ to be a multiple of $2$,
but this is easily achieved by locally multiplying $a$ and $d$ with $2$,
and performing the division using $a^\prime = 2 a$ and $d^\prime = 2 d$.
Note that the output of the division remains unchanged by~this.

For the remaining error term
$e = \frac{r_a+qr_p - (r_1+r_2)}{d}$,
each value $r_a, r_p$ and $r_j$ is less than $d$, and hence, $-1 \le e\le 2$.
In our protocols, we reduce this error to $0 \le e \le 2$ 
by adding a combination of $\tash{q}$ and appropriate constants to the output.

\subsubsection{Passively Secure Protocols}
We propose passively secure division protocols
in Protocol~\ref{alg:division} and \ref{alg:divisionLSS}.
The first protocol works for input $\tash{a}$, where $a$ is a multiple of $2$,
and the second for $\pash{a}$ by extending the first protocol.
We further extend our division protocols to \emph{signed} integers $a$ in Appendix~\ref{sect:divisionsigned}.

Both Protocol~\ref{alg:division} and \ref{alg:divisionLSS}  have probabilistic rounding
that outputs $a/d$, $a/d+1$, or $a/d + 2$.
In other words, our protocols guarantees that there is only a small difference between $\frac{a}{d}$ and 
the output of our protocols,
while the standard approach to division protocols have a similar rounding difference \emph{and} 
a large error $q\alpha_p$ with a certain probability.
The functionalities of these protocols appear in Appendix~\ref{sect:functionality}.
Note, the most interesting case in which $p$ is a Mersenne prime and $d$ is a power of 2,
the protocol outputs either $a/d$ or $a/d + 1$.

\begin{algorithm}[t] \small
\caption{Secure Division by Public Value in $\tash{\cdot}$}
 \label{alg:division}
 \begin{algorithmic}[1]
  \Functionality $\tash{c} \gets \Div_{(2,2)}(\tash{a}, d)$
  \Require Share of dividend $\tash{a}$ and (public) divisor $d$, where $a$ and $d$ are even numbers.
  \Ensure $\tash{c}$, where $c \approx \frac{a}{d}$.
  \State Let $\alpha_p$ and $r_p$ be $p=\alpha_p d + r_p$, where $0 \le r_p < d$.
  \State $\tash{q} \gets \Fqt(\tash{a})$
  \State $P_1$ computes $\tash{b}_1 \gets \intdiv{\left(\tash{a}_1 + d-1 -r_p\right)}{d}$.
      \Comment{``in $\NN$'' means no reduction $\mod p$}
  \State $P_2$ computes $\tash{b}_2 \gets \intdiv{\tash{a}_2}{d}$ in $\NN$
  \State $\tash{c} := \tash{b} -(\alpha_p+1)\tash{q}+ 1$
 \end{algorithmic}
\end{algorithm}


\begin{algorithm}[t]\small 
\caption{Secure Division by Public Value in $\pash{\cdot}$}
 \label{alg:divisionLSS}
 \begin{algorithmic}[1]
  \Functionality $\pash{c}\gets \Div_{(2,3)}(\pash{a}, d)$
  \Require Share of dividend $\pash{a}$ and public divisor $d$, where $0 \le a \le 2^{\left|p\right|-1}-1$
  \Ensure $\pash{c}$, where $c \approx \frac{a}{d}$.
  \State $\tash{a}\gets \ptotconvert(\pash{a})$ 
  \State $\tash{a^\prime} := \tash{2a}, d^\prime := 2d$
  \State $\tash{c}\gets \Div_{(2,2)}(\tash{a^\prime}, d^\prime)$
  \State $\pash{c} \gets \ttopconvert(\tash{c})$
  \State Output $\pash{c}$
 \end{algorithmic}
\end{algorithm}

\subsubsection{Output of Protocols}

Consider the reconstruction of the output shares calculated returned by Protocol~\ref{alg:division}. We obtain (see Section~\ref{sec:intuition} for notation):
{ \small
\begin{align*}
& \left(\tash{b}_1-(\alpha_p +1)\tash{q}_1+ 1 \right) + \left(\tash{b}_2-(\alpha_p +1)\tash{q}_2\right) \\
&= \tash{b}_1 + \tash{b}_2 - (\alpha_p+1)(\tash{q}_1+\tash{q}_2) +1 \\
&= (\tash{a}_1 + d - 1 -r_p)/d + \tash{a}_2/d - (\alpha_p+1)q + 1 \\
&= (\alpha_1 d + r_1+d-1-r_p)/d + (\alpha_2 d + r_2)/d - q\alpha_p-q + 1\\
&= \alpha_1 + \alpha_2 - q\alpha_p-q + 1 + (r_1-r_p +d-1)/d + r_2/d 
\end{align*}
}
From Eq.~(\ref{eq:simpledivision}) and the fact that $r_1 \le d-1$, the last equation equals to 
\begin{equation}\label{eq:division}\small
    \alpha_a  -q + 1 + \frac{r_a + qr_p - r_1 -r_2}{d} + (r_1-r_p +d-1)/d.
\end{equation}

The above computation shows the term $q\alpha_p$ being eliminated.
The remaining terms are quite small
since 
$q \in \bit$, each $r_i$ is smaller than $d$, and $\frac{r_a + qr_p - r_1 -r_2}{d}$ and $(r_1-r_p +d-1)/d$ are at most 1.
To see that the output in  Eq.~(\ref{eq:division}) corresponds to the output defined in the ideal functionalities defined in Appendix~\ref{sect:functionality}, a more precise analysis is required.
In Appendix~\ref{sect:divgeneral} we provide a detailed analysis of both the simpler specific case of $p$ being a Mersenne prime and $d$ a power of $2$, and the general case.
Note that this analysis is required to formally establish the security of our protocols.

Lastly, note that the distribution of the output expressed in Eq.~(\ref{eq:division}), depends on the value of the input \textit{shares} (as opposed to the reconstructed value).
We provide a full analysis of the output distribution in Appendix~\ref{sect:divprob}.
%
%
The output range and distribution of  Protocol~\ref{alg:divisionLSS} follows that of  Protocol~\ref{alg:division}.


\subsubsection{Security}
The following theorem establishes security of 
Protocol~\ref{alg:division}.
\begin{thm}
Protocol~\ref{alg:division} securely computes the division functionality $\Fdiv$ 
in the $\Fqt$-hybrid model in the presence of a passive adversary.
\end{thm} 

The proof of this immediately follows from the correctness discussion in above,
since Protocol~\ref{alg:division} only consists of local computations except for $\Fqt$.

The security of Protocol~\ref{alg:divisionLSS} follows from that of Protocol~\ref{alg:division}.
Here, $\ttopconvert$ is the only additional step requiring communication compared to Protocol~\ref{alg:division}.
This step can easily be simulated since the output consists only of shares,
and the simulator can simply insert a random share as to simulate transcripts.

\subsubsection{Efficiency}
We obtain the concrete efficiency of our protocols by considering efficient instantiations of the required building blocks. 

The quotient transfer protocol in \cite{ACISP:KIMHC18} requires $2$ bits of communication and $1$ communication round,
besides a single call of $\Fmodconv$.
The modulus conversion protocol in \cite{ACISP:KIMHC18} and $\ttopconvert$ 
require $3|p|+3$ bits and $2|p|$ bits of communication, respectively, and both $1$ communication round.
Furthermore, we can reduce the number of rounds required in Protocol \ref{alg:divisionLSS}
by parallel execution\footnote{We change the protocol by 
skipping $\ptotconvert$ in step 4 in Protocol~\ref{alg:qt} (hence the output of $\Fqt$ is $\pash{q}$)
and local computation in step 5 in Protocol~\ref{alg:division} 
is performed over $\pash{\cdot}$-sharings by computing $\ttopconvert(\tash{b})$ just before this step.
} of $\Fqt$ and $\ttopconvert$.
Consequently,
instantiating $\Fqt$ (and $\Fmodconv$ used in $\QT$ internally) by the protocols in \cite{ACISP:KIMHC18},
Protocol~\ref{alg:division} and \ref{alg:divisionLSS} require $3|p| + 5$ and $5|p| + 5$ bits of communication and $2$ communications rounds in total.

\subsubsection{Comparison}
We compare Protocol~\ref{alg:divisionLSS} with the one (denoted truncation in the original paper) of  ABY3~\cite{CCS:MohRin18}, which has been used in subsequent literature, such as FALCON.
Since the ABY3 protocol is proposed as truncation, we let $d=2^\delta$.
The ABY3 protocol requires $6(2|p|-\delta -1)$ bits and $|p| -1$ rounds in the offline phase, and
$3|p|$ bits and 1 round in the online phase, 
so the total communication cost is $15|p|-6\delta-6$ bits and $|p|$ rounds.
In comparison, our protocol requires $5|p|+5$ bits and $2$ rounds.
Thus, the total cost of our protocol is much better compared to ABY3's,
and slightly worse than ABY3's \emph{online} cost.
In addition, the ABY3's protocol can contain a large error $\frac{p}{d}$ with a certain probability.
This leads to a small message spaces and/or less accuracy, as pointed out in Sec.~\ref{sec:currentdiverror}.

\subsubsection{Active Security}
In the above, we have only treated passively secure protocols.
We next outline how we construct a division protocol satisfying active security with abort.
A quotient transfer protocol secure against an active adversary with abort has already been proposed in \cite{ACISP:KIMHC18}.
We can thus employ the same approach as in the passive security:
eliminating $q\alpha_p$ via the use of $\Fqt$.

The difference between our actively and passively secure division protocols, is
that the only known efficient actively secure quotient transfer protocol uses input 
$\pash{a}$, where $a$ is required to be a multiple of 4.
Hence, our actively secure division protocol works
directly on $\pash{a}$ and essentially executes the following steps:
\begin{enumerate}
    \item $\pash{a'} := 4\pash{a}$ and $d' := 4d$
    \item $\pash{q} \gets \Fqt(\pash{a})$
    \item Parties divide own sub-shares of $\pash{a'}$ by $d'$, and let them be $\pash{b}$.
    \item $\pash{c} := \pash{b} - \alpha_p \pash{q}$
\end{enumerate}
In addition, we further adjust the output using appropriate constants to minimize the difference between $\frac{a}{d}$ and the protocol output.
The concrete protocol appears in Appendix~\ref{sect:divisionmal}.

Regarding security, as we do not convert to $\tash{a}$,
the quotient transfer protocol is the only step that requires communication,
and the security of the protocol is thus trivially reduced to~$\Fqt$.

\section{Elementary functions for machine learning}
\label{section:elementary}
In the following, we will present efficient and high-accuracy protocols for arithmetic functions suitable for machine learning, such as inversion, square root extraction, and exponential function evaluation. 
All of these rely on fixed-point arithmetic, and will make use of the division protocols introduced above to implement this. 
Recall that, the product of fixed-point values $a$ and $b$ with offset $\ell$, written $a \times_{\ell} b$, corresponds to (standard) multiplication of $a$ and $b$, followed by division by $2^\ell$.
To ease notation in the protocols, we use $\pash{a} \cdot \pash{b}$ to denote $\Fmult(\pash{a},\pash{b})$, 
and $\pash{a} \times_\ell \pash{b}$ to denote $\Fdiv(\pash{a} \cdot \pash{b}, 2^\ell)$. 

All protocols will explicitly have as parameters the offset of both input and output values, often denoted $\alpha$ and $\delta$, and in particular will allow these to be different.
This can be exploited to obtain more accurate computations when reasonable bounds for the input and output are publicly known.
For example, consider the softmax function, often used in neural networks,
defined as 
$\frac{e^{u_i}}{\sum_{j = 0}^{k-1}e^{u_j}} = \frac{1}{\sum_{j = 0}^{k-1}e^{u_j-u_i}}$ for input $(u_0,\ldots,u_{k-1})$.
The output is a value between 0 and 1, and to maintain high accuracy, the offset should be large, e.g. $23$ to maintain $23$ bits of accuracy below the decimal point.
However, the computation of $\sum_{j = 0}^{k-1}e^{u_j-u_i}$ is often expected to be a large value in comparison, and a much smaller offset can be used to prevent overflow e.g. $-4$.
Furthermore, we highlight that internally, some of the protocols will switch to using an offset different from $\alpha$ and $\delta$ to obtain more accurate numerical computations.
By fine-tuning and tailoring the offsets to the computations being done, the most accurate computation with the available $\ell$ bits for shared values, can be obtained. 


We will first show protocols for unsigned inputs, and defer the extension to signed inputs to Sect.~\ref{sect:elementarysigned}.
\jacob{Another option is to highlight this after each description of the relevant protocols.}

\subsection{Inversion}
In the following, we introduce a protocol for computing the inverse of a shared fixed-point value. This is a basic operation required for many computations, including machine learning.

Before presenting the inversion protocol itself, we introduce a specialized bit-level functionality of \emph{private scaling} that will compute a representation of the input $\pash{a}$ which allows us to make full use of the available bit range for shared values. Specifically, the representation of $\pash{a}$ is $\pash{b} = \pash{a} \cdot \pash{c}$, where $2^{\ell -1} \leq b \leq 2^{\ell} - 1$ and $c$ is a power of $2$ (recall that shared values are $\ell$ bit integers).
This functionality corresponds to a left-shift of the shared value $\pash{a}$ such that the most significant \emph{non-zero} bit becomes the most significant bit, where $c$ represents the required shift to obtain this.
We will denote this operation $\MSNZBFit$ (MSNZB denoting Most Significant Non-Zero Bit) and the corresponding functionality $\Fmsnzbfit$. 
The protocol presented in Protocol~\ref{alg:msnzbfitting} implements this functionality.
Recall that $\Fbitcomp$ and $\Fbitdecomp$ are the functionalities of bit-(de)composition.

\begin{thm} 
Protocol~\ref{alg:msnzbfitting} securely implements $\Fmsnzbfit$ in the ($\Fbitdecomp$, $\Fbitcomp$, $\Fmult$)-hybrid model in the presence of a passive adversaries. 
\end{thm}

\begin{algorithm}[ht] \small
\caption{MSNZB Fitting}
 \label{alg:msnzbfitting}
 \begin{algorithmic}[1]
  \Functionality $(\pash{b}, \pash{c})\gets \MSNZBFit(\pash{a})$
  \Require $\pash{a}$
  \Ensure $\pash{b}, \pash{c}$, where $\pash{b} = \pash{a} \cdot \pash{c}$, $2^{\ell -1} \leq b \leq 2^{\ell} - 1$, and $c = 2^e$ for some $e \in \mathbb{N}$.
  \Parameter $\ell$
  \State $(\bash{a_1},\dots,\bash{a_{\ell}}) \gets \Fbdc(\pash{a})$
  \State $\bash{f_{\ell}}:=\bash{a_{\ell}}$
  \For{$i = \ell-1$ \text{ to } $1$}
  \State $\bash{f_i}:=\bash{f_{i+1}}\vee \bash{a_i}$ \Comment{$f_i = 1$ for all $i$ corresponding to MSNZB of $a$ or smaller}
  \EndFor
  \State $\bash{x_{\ell}}:=\bash{a_{\ell}}$
  \For{$i = \ell-1$ \text{ to } $1$ }
  \State $\bash{x_i}:= \bash{f_i} \oplus \bash{f_{i+1}}$ \Comment{$x_i = 1$ only for $i$ corresponding to MSNZB of $a$}
  \EndFor
  \State $\pash{c} \gets \Fbitcomp(\bash{x_{\ell}},\dots, \bash{x_1})$
  \Comment{Bit-compose $\bash{x_i}$ \emph{in the reverse order} to obtain $c = 2^{\ell-1-\floor{\log_2{a}}}$}
  \State $\pash{b} = \pash{a} \cdot \pash{c}$ 
  \State Output $\pash{b}$ and $\pash{c}$
 \end{algorithmic}
\end{algorithm}

Using $\MSNZBFit$ as a building block, we now construct our inversion protocol.
The protocol is based on the Taylor series for $(1-x_0)^{-1}$ centered around $0$, where $x\in [0; \frac{1}{2})$:
\begin{equation}\label{eq:taylor}\small
    \frac{1}{1-x_1} = \sum_{i = 0}^\infty x_1^i = 1 + x_1 + x_1^2 + \cdots.
\end{equation}
Continuing this Taylor series until the $n$-degree,
yields the remainder term $\frac{x^{n+1}}{1-x} \le \frac{1}{2^n}$, which implies that the approximation has $n$ bits accuracy.

Firstly, we use $\MSNZBFit$ to left-shift input $\pash{a}$ to obtain  $\pash{b} = \pash{a} \cdot \pash{c}$.
Interpreting the resulting value $\pash{b}$ as being a fixed-point value with offset $\ell$ implies that $b \in [\frac{1}{2};1)$.
This representation forms the basis of our computation.
(Note that since $b \in [\frac{1}{2};1)$, we have that $\frac{1}{b} \leq 2$, ensuring that $\frac{1}{b}$ can be represented using $\ell +1$ bits.)

For the computation of $\frac{1}{b}$, instead of using Eq.~\eqref{eq:taylor} directly, which requires $r$ multiplications for $(r+1)$-th degree terms, we use the following product requiring only $\log r$ multiplications. 
\begin{equation}\label{eq:prodtaylor}\small
  \prod_{j=0}^\infty (1+x_1^{2^j}) = (1+x_1)(1 + x_1^2)(1 + x_1^4)\cdots.
\end{equation}
Letting $b = 1 - x_1$ (which ensures $x_1 \in [0;\frac{1}{2})$), our inversion protocol shown in Protocol~\ref{alg:inversion} iteratively computes Eq.~(\ref{eq:prodtaylor})
by first setting $x_1 = 1-b$ and $y_1 = 1+x_1 = 2-b$ (Step $2$-$3$), 
and in each iteration computing $(1 + x_1^i)$ and multiplying this with $y_i$ (Step $4$-$6$).
The number of iterations is specified via the parameter $I$.
Finally, to obtain (an approximation to) $\pash{\frac{1}{a}}$, 
we essentially only need to scale the computed $\pash{y_I} = \pash{\frac{1}{b}}$ with $\pash{c}$ (as $\frac{1}{b} \cdot c = \frac{1}{ac} \cdot c = \frac{1}{a}$).
Note, however, that the output has to be scaled taking into account the input and output offsets, as well as the offset used in the internal computation. 
To see that the correct scaling factor is $2^{\alpha + \delta - 2 \ell}$, note that for input $a = a'_{\pre{\alpha}}$ and output $b = a \cdot c = b'_{\pre{\ell}}$ of $\MSNZBFit$, we have
\[\small
y_I = \frac{2^\ell}{b'} = \frac{2^\ell}{a' \cdot 2^\alpha \cdot c \cdot 2^{-\ell}} = \frac{1}{a' \cdot c} \cdot 2^{2 \ell - \alpha}
\] 
and that the output should be scaled with $2^\delta$.

We define the corresponding functionality $\Finvers$
in which on input of shares computes the above Taylor series expansion and output shares of that output.

%
%

\begin{algorithm}[t]\small 
\caption{Inversion}
 \label{alg:inversion}
 \begin{algorithmic}[1]
  \Functionality $\pash{d} \gets \Inv(\pash{a})$ 
  \Require $\pash{a}$, where $a \in \widehat{\mathbb{Q}}_{\langle \ell, \alpha \rangle}$ 
  \Ensure $\pash{d}$, where $d \approx \left(\frac{1}{a}\right)_{\pre{\delta}}$
  \Parameter$(\ell, I, \alpha, \delta)$, where $I$ is the number of iterations (say, $I=\ceil{\log{\ell}}$) used in the computation
  \State  
  $(\pash{b}, \pash{c})\gets \MSNZBFit(\pash{a})$ \Comment{$b = b' \cdot 2^\ell$ where $b' \in[\frac{1}{2},1)$} 
  \State $\pash{x_1} := 1_{\pre{\ell}} - \pash{b}$ 
  \State $\pash{y_1} := 2_{\pre{\ell}} - \pash{b}$
  \For{$i = 2 \text{ to } I$}
  \State
  $\pash{x_i}:=\pash{x_{i-1}} \times_{\ell} \pash{x_{i-1}}$
    \State  
  $\pash{y_i}:=\pash{y_{i-1}} \times_{\ell} ( 1_{\pre{\ell}} + \pash{x_{i-1}})$
  \EndFor
 \State Output $\pash{y_{I}} \cdot \pash{c} \cdot 2^{\alpha + \delta - 2 \ell}$ 
 \end{algorithmic}
\end{algorithm}

\begin{thm}
The protocol $\Inv$ securely computes inversion functionality $\Finvers$ 
in the ($\Fmsnzbfit, \Fdiv, \Fmult$)-hybrid model
in the presence of a passive adversary.
\end{thm}

\subsection{Division with Private Divisor}
Given our protocol for inversion, it becomes trivial to construct a high accuracy protocol for division with a private divisor.
Specifically, for values $\pash{a}$ and $\pash{d}$, we simply compute $\pash{\frac{1}{d}}$ using $\Inv$, and multiply this with $\pash{a}$ to obtain $\pash{\frac{a}{d}}$.
The resulting protocol, $\Divpriv$, is shown in Protocol~\ref{alg:divpriv}.
Note that the accuracy of the result is determined by the parameter $I$ of the inversion protocol.
Setting $I = \log \ell$ gives $\ell$ bits of precision for the inversion, which ensures the result is equal to $\frac{a}{d}$ for $\ell$-bit fixed-point values. 
We define the corresponding functionality $\Fdivpriv$ that 
as on input $\pash{a}$ and $\pash{d}$ outputs $\pash{\frac{a}{d}}$
in which $\pash{\frac{1}{d}}$ is obtained by $\Finvers$.

 \begin{thm}
 The protocol $\Divpriv$ securely computes fixed-point division $\Fdivpriv$
 in the ($\Finvers, \Fdiv, \Fmult$)-hybrid model
 in the presence of a passive adversary.
 \end{thm}

%
%

\begin{algorithm}[t]\small 
\caption{Integer Division with Private Divisor}
 \label{alg:divpriv}
 \begin{algorithmic}[1]
  \Functionality $\pash{z} \gets \Divpriv(\pash{a},\pash{d})$ 
  \Require $\pash{a},\pash{d}$, where $a = a'_{\pre{\alpha}} \in \widehat{\mathbb{Q}}_{\langle \ell, \alpha \rangle}$, and $d = d'_{\pre{\beta}} \in \widehat{\mathbb{Q}}_{\langle \ell, \beta \rangle}$
  \Ensure $\pash{z}$, where $z \approx \left(\frac{a'}{d'}\right)_{\pre{\delta}}$
  \Parameter $(\ell,I,\alpha,\beta,\delta)$ where $\alpha$ and $\beta$ are the offsets of $a$ and $d$, respectively.
  \State 
  $\pash{z'} \gets \Inv_{(\ell,I,\beta,\delta)}(\pash{d})$ \Comment{$z' = \left(\frac{1}{d'}\right)_{\pre{\delta}}$}
  \State Output $\pash{z'} \cdot \pash{a} \cdot 2^{-\alpha}$
  \end{algorithmic}
\end{algorithm}

%
%

\subsection{Square Root and Inverse Square Root}
Computing the inverse of the square root of an input value, is a useful operation for many computations, \eg normalization of a vector, and is likewise used in Adam.
Hence, having an efficient protocol for directly computing this, is beneficial.

Our protocol for computing the inverse of a square root is shown in Protocol~\ref{alg:invsqrt}, and is based on Newton's method for the function $f(y) = \frac{1}{y^2} - x$ for input value $x$ (note that $f(y') = 0$ implies $y' = \frac{1}{\sqrt{x}}$).
This involves iteratively computing approximations
\[\small
y_{n+1} = y_n - \frac{f(y_n)}{f'(y_n)} = \frac{y_n(3 - x \cdot y^2)}{2} 
\]
for an appropriate initial guess $y_0$ (Step $4$-$5$ performs this iteration).
To ensure fast convergence for a large range of input values, we represent the input $x = b \cdot 2^e$ for $b \in [\frac{1}{2};1)$, which implies \[\small
\frac{1}{\sqrt{x}} = 
\begin{cases} 
	\left(\frac{1}{\sqrt{b}}\right) \cdot 2^{-e/2} & \text{if $e$ is even} \\ 
	\left(\frac{\sqrt{2}}{\sqrt{b}}\right) \cdot 2^{-(e+1)/2} & \text{if $e$ is odd}
\end{cases}.
\] 
Hence, we only need to compute $\frac{1}{\sqrt{b}}$ for $b \in [\frac{1}{2};1)$, in which case using $1$ as the initial guess provides fast convergence.
However, the parties should not learn which of the above two cases the input falls into.
We introduce a sub-protocol, $\MSNZBFitExt$ shown in Protocol~\ref{alg:subprotocol}, that computes values $r$ and $c'$, where $(r,c') = (0,2^{e/2})$ if $x$ falls into the first case, and $(r,c') = (1,2^{(e+1)/2})$ otherwise.
Note that like $\MSNZBFit$, the extended $\MSNZBFitExt$ right-shifts the input $a$ to obtain an $\ell$-bit value $b = a \cdot c$, that when interpreted as an element of $\widehat{\mathbb{Q}}_{\langle \ell, \ell \rangle}$, represents $b \in [\frac{1}{2};1)$, and that $c' = \sqrt{c}$. 
Having $r$ and $c'$ allows us to compute $\frac{1}{\sqrt{x}}$ as $\frac{1}{\sqrt{b}} \cdot (1 + r \cdot (\sqrt{2} - 1)) \cdot c'$.
Finally, note that the output has to be scaled, taking into account the input and output offsets, as well as the offset used for the internal computation. 
To see that the correct scaling factor is $2^{\delta - \frac{3}{2} \ell + \frac{\alpha}{2}}$, note that for input $a = a'_{\pre{\alpha}}$ and output $b = a \cdot (c')^2 = b' * 2^\ell$ of $\MSNZBFitExt$, we have
\[\small
y_I \approx \frac{2^\ell}{\sqrt{b'}} = \frac{2^\ell}{\sqrt{a' \cdot 2^\alpha \cdot (c')^2 \cdot 2^{-\ell}}} = \frac{1}{\sqrt{a'}} \cdot \frac{1}{c'} \cdot 2^{\ell/2 - \alpha / 2 }
\]
and that the output should be scaled with $2^\delta$. 
We define the functionality $\Finvsqrt$ that computes $\frac{1}{\sqrt{x}}$ as done in the above
using the Newton's method.

\begin{algorithm}[t]\small 
\caption{$\MSNZBFitExt$ Sub-protocol for $\InvSqrt$}
 \label{alg:subprotocol}
 \begin{algorithmic}[1]
  \Functionality 
  $(\pash{b}, \pash{c^\prime},\pash{r})\gets \MSNZBFitExt(\pash{a})
  $ 
  \Require $\pash{a}$
  \Ensure $(\pash{b}, \pash{c^\prime},\pash{r})$, 
  where $b = b'_{\pre{\ell}} \in \widehat{\mathbb{Q}}_{\langle \ell, \ell \rangle}$ and  $b' \in [\frac{1}{2};1)$, $x = b' \cdot 2^e$, and $r =0$ if $e$ is even, and $r = 1$ otherwise.
  \Parameter $\ell$
  \State Parties jointly execute steps $1$-$9$ of protocol $\MSNZBFit$.
  \State
  $\ell^\prime := \floor{\frac{\ell}{2}}$
  \State $\bash{x_i^\prime}:=\bash{x_{\ell+1-i}}$ for $1 \le i \le \ell$
  \For{$i = 1 \text{ to } \ell^\prime-1$}
  \State $\bash{y_i}:=\bash{x_{2i}^\prime}\oplus\bash{x_{2i+1}^\prime}$
  \EndFor
  \If{$\ell$ is an even number}
    \State $\bash{y_{\ell'}}:=\bash{x_{2\ell'}^\prime}\oplus\bash{x_{2\ell'+1}^\prime}$
    \Else
    \State $\bash{y_{\ell^\prime}}:= \bash{x_{2\ell^\prime}^\prime}$
  \EndIf
  \State 
  $\bash{r}:= \bash{x_{2}^\prime} \oplus \bash{x_{4}^\prime} \oplus \dots \oplus \bash{x_{2\floor{\frac{\ell'}{2}}}^\prime}$ 
  \State $\pash{r} \gets \Fmodconv(\bash{r})$
  \State $\pash{c^\prime}\gets\Fbitcomp(\bash{y_1}, \dots, \bash{y_{\ell^\prime}})$
 \State Output $(\pash{b}, \pash{c^\prime},\pash{r})$
 \end{algorithmic}
\end{algorithm}

%
%

\begin{algorithm}[t]\small 
\caption{Inversion of Square Root}
 \label{alg:invsqrt}
 \begin{algorithmic}[1]
  \Functionality $\pash{z} \gets \InvSqrt(\pash{a})$ 
  \Require $\pash{a}$, where $a = a'_{\pre{\alpha}} \in \widehat{\mathbb{Q}}_{\langle \ell, \alpha \rangle}$
  \Ensure $\pash{z}$, where $ z \approx \left(\frac{1}{\sqrt{a'}}\right)_{\pre{\delta}}$
  \Parameter$(\ell, I, \alpha, \delta)$, where $I$ is the number of iteration (say, $I=\ceil{\log{\ell}}$) in the computation. 
  \State $(\pash{b},\pash{c'},\pash{r}) \gets \MSNZBFitExt(\pash{a})$
  \State $\pash{x_1} := 3_{\pre{\ell}} - \pash{b}$
  \State $\pash{y_1} := \pash{x_1} / 2$
  \For{$i = 2 \text{ to } I$}
  \State 
  $\pash{x_i} :=3_{\pre{\ell}} - (\pash{y_{i-1}} \times_{\ell} \pash{y_{i-1}}) \cdot \pash{b}$
    \State 
  $\pash{y_i} :=\pash{x_{i-1}} \times_{\ell+1} \pash{y_{i-1}}$ \Comment{Implicit scaling by $\frac{1}{2}$}
  \EndFor
 \State Output $\pash{y_{I}} \cdot (1 + \pash{r} \cdot (\sqrt{2} - 1)) \cdot \pash{c^\prime} \cdot 2^{\delta - \frac{3}{2} \ell + \frac{\alpha}{2}}$ 
 \end{algorithmic}
\end{algorithm}

\begin{thm}
The protocol $\InvSqrt$ securely computes the inverse of the square root functionality $\Finvsqrt$ 
in the ($\Fmsnzbfit,\Fmodconv,\Fdiv,\Fmult$)-hybrid model
in the presence of a passive adversary.
\end{thm}

Given the above protocol $\InvSqrt$ for computing $\frac{1}{\sqrt{x}}$, and noting that $\sqrt{x} = \frac{x}{\sqrt{x}}$, we can easily construct a protocol for computing $\sqrt{x}$, simply by running $\InvSqrt$ and multiplying the result with $x$.
The resulting protocol, $\Sqrt$, is shown in Protocol~\ref{alg:sqrt}.
Let $\Fsqrt$ be the functionality 
that on input $\pash{a}$ outputs $\pash{\sqrt{a}}$
in which $\pash{\frac{1}{\sqrt{a}}}$ is obtained by $\Finvsqrt$.

\begin{thm}
The protocol $\Sqrt$ securely computes the square root functionality $\Fsqrt$ 
in the ($\Finvsqrt,\Fdiv,\Fmult$)-hybrid model
in the presence of a passive adversary.
\end{thm}

%
%

\begin{algorithm}[t]\small 
\caption{Square Root}
 \label{alg:sqrt}
 \begin{algorithmic}[1]
  \Functionality $\pash{z} \gets \Sqrt(\pash{a})$ 
  \Require $\pash{a}$, where $a = a'_{\pre{\alpha}} \in \widehat{\mathbb{Q}}_{\langle \ell, \alpha \rangle}$
  \Ensure $\pash{z}$, where $z \approx (\sqrt{a'})_{\pre{\delta}}$
  \Parameter $(\ell,I,\alpha,\delta)$ where $I$ is the number of iterations used in the computation.
  \State $\pash{z'} \gets \InvSqrt_{(\ell,I,\alpha,\delta)}(\pash{a})$ \Comment{$z' = \left(\frac{1}{\sqrt{a'}}\right)_{\pre{\delta}}$} 
 \State Output $\pash{a} \cdot \pash{z} \cdot 2^{-\alpha}$
 \end{algorithmic}
\end{algorithm}

\subsection{Exponential Function}
To obtain a fast and highly accurate protocol for evaluating the exponential function, we adopt 
what we call \emph{hybrid table-lookup/series-expansion} technique. Intuitively, it utilizes the table lookup approach for the large-value part of the input, in combination with the Taylor series evaluation for its small-value counterpart.
%
%
We first recall that the Taylor series of the exponential function is
\begin{align*}\small
    \exp{x} = \sum_{i = 0}^\infty \frac{x^i}{i!} = 1 + x + \frac{x^2}{2} +  \frac{x^3}{6} +  \frac{x^4}{24} + \cdots.
\end{align*}
which converges fast for small values of $x$.
To minimize the value for which we use the above Taylor series, we separate the input $a$ into three parts: 
\begin{enumerate}\setlength{\leftskip}{-2mm}
    \item $\mu$: a lower bound for the input
    \item $b_\ell,\dots,b_{\ell-t}$: bit representation of the $t$ most significant bits of $b:=a-\mu$
    \item $b_\sigma$: integer representation of $(a-\mu) - \sum_{\ell-t \le i\le \ell}2^{i-\alpha} b_i$
\end{enumerate}
which means that we can compute $\exp (a)$ as 
\begin{equation}\small
 \exp a = \left( \prod_{i=\ell-t}^\ell \exp (b_i \cdot 2^{i-\alpha})\right) \cdot \exp(b_\sigma) \cdot \exp (\mu).
\end{equation}
Here, $\alpha$ is the input offset and $t$ is a parameter of our protocol that determines which part of the input we will evaluate using table lookups, and which part we will evaluate using a Taylor series.
In this product, we evaluate $\exp(\mu)$ locally (as $\mu$ is public),
$\prod_{i=\ell-t}^\ell \exp (b_i \cdot 2^i)$ using table lookups,
and $\exp(b_\sigma)$ using the Taylor series. 
The taylor series rapidly converge since $b_\sigma$ is made small
due to the subtraction of $\mu$ and the value of the $t$ most significant bits of $a$.

More specifically, for the table lookup computation, note that the binary value $b_i$ determines whether the factor $\exp 2^i$ will included.
Hence, by combining bit decomposition, that allows parties to compute $\bash{b_i}$ from $\pash{b}$, with the $\mathsf{CondAssign}$ protocol using $\bash{b_i}$ as the condition, and the values $1$ and $\exp 2^i$, which are public and can be precomputed, 
we obtain an efficient mechanism for computing $\prod_{i=\ell-t}^\ell \exp (b_i \cdot 2^i)$.
However, to maintain high accuracy, the parties will not use $\exp 2^i$ directly, but precompute a mantissa $f_i$ and exponent $2^{\epsilon_i}$ such that $f_i \cdot 2^{\epsilon_i} = \exp 2^{i-\alpha}$.
This allows the parties to compute the product of $f_i$ values separately from the product of $2^{\epsilon_i}$ values, and only combine these in the final step constructing the output, 
 thereby avoiding many of the rounding errors that potentially occur in large products of increasingly larger values. 

Lastly, the result is computed as $\prod_i f_i \cdot \prod_i 2^{\epsilon_i} \cdot \exp a_\sigma$.
Note that the input and output offsets have to be taken into account, and the output adjusted appropriately.
We define the functionality $\Fexp$ such that 
on input $\pash{a}$, $e^a$ is computed as done in the above
and output $\pash{e^a}$.

\begin{algorithm}[t]\small 
\caption{Exponential Function}
 \label{alg:exponent}
 \begin{algorithmic}[1]
  \Functionality $\pash{z} \gets \Exponent(\pash{a})$ 
  \Require $\pash{a}$, where $a = a'_{\pre{\alpha}} \in \widehat{\mathbb{Q}}_{\langle \ell, \alpha \rangle}$
  \Ensure $\pash{z}$, where $z \approx (\exp a')_{\pre{\delta}}$
  \Parameter $(\ell,I, \alpha, \beta, \delta, \mu, t)$ where $I$ is the number of iterations used in the computation, $\beta$ is the offset of the lookup table values, $t$ indicates the lookup table vs Taylor series threshold, and $\mu$ is a lower bound for the input.
  \State $\pash{b} := \pash{a} - \mu_{\pre{\alpha}}$
  \State 
  $\bash{b_{\ell}}, \dots, \bash{b_{\ell-t}} \gets \Fbdc(\pash{b})$ \Comment{We use only $\ell - t$ MSBs while $\Fbitdecomp$ outputs $\ell$ bits.}
  \State $\pash{b_i} \gets \Fmodconv(\bash{b_i})$ for $i = \ell,\ldots \ell-t$.
  \State $\pash{b_\sigma}:=\pash{b} - \displaystyle\sum_{\ell-t \le i\le \ell}2^{i}\pash{b_i}$ \Comment{Value of $t$ LSBs of $\pash{b}$}
  \State Parties define $f_i,\epsilon_i$ such that $\exp 2^{i - \alpha} = f_i \cdot 2^{\epsilon_i}$ \Comment{Precomputed lookup table values}
  \State Using $\pash{f_i^\prime} \gets \Fcondassign(1_{\pre{\beta}}, (f_i)_{\pre{\beta}},\bash{b_i})$, the parties obtain
  $\pash{f_i^\prime} =
  \begin{cases}
  (f_i)_{\pre{\beta}}&\text{if }b_i=1\\
  1_{\pre{\beta}}&\text{otherwise}
  \end{cases}$
   for $i=\ell,\ldots,\ell -t$
  \State Using $\pash{\epsilon_i^\prime} \gets \Fcondassign(1,2^{\epsilon_i},\bash{b_i})$, the parties obtain
  $\pash{\epsilon_i^\prime}=
  \begin{cases}
  2^{\epsilon_i}&\text{if }b_i = 1\\
  1&\text{otherwise}
  \end{cases}$
for $i=\ell,\ldots,\ell -t$
  \State $\pash{f^\prime}:=\pash{f_{\ell}^\prime} \times_\beta \ldots \times_\beta  \pash{f_{\ell - t}^\prime}$
    \State $\pash{\epsilon^\prime}:= \pash{\epsilon_\ell^\prime} \cdot \ldots \cdot \pash{\epsilon_{\ell -t}^\prime}$
  \State $b_{\sigma,0} := 1$
  \For{$i=1 \text{ to } I-1$} 
  \State $\pash{b_{\sigma,i}} \gets \pash{b_{\sigma,i-1}} \times_{\alpha} \pash{b_\sigma}$ 
  \EndFor 
   \State $\pash{b_\sigma^\prime}:=\displaystyle\sum_{0\le i<I}\frac{\pash{b_{\sigma,i}}}{i!}$ \Comment{Division using $\Div$}
  \State Output $\pash{f^\prime} \cdot \pash{\epsilon^\prime} \cdot \pash{b_\sigma^\prime} \cdot \exp \mu \cdot 2^{\delta -\beta -\alpha}$ 
 \end{algorithmic}
\end{algorithm}

\begin{thm}
The protocol $\Exponent$ securely computes the exponential function functionality $\Fexp$ 
in the ($\Fcondassign,\Fmodconv,\Fbdc,\allowbreak \Fdiv,\Fmult$)-hybrid model
in the presence of a passive adversary.
\end{thm}

\subsection{Extension to signed integer}\label{sect:elementarysigned}
We extend the protocols proposed in this section 
into those for signed integers.
We generically convert the protocols by extracting sign and absolute of an input at first.
The protocol that extract sign and absolute appears in Appendix~\ref{seq:signed}.
Obtaining sign and absolute, 
we conduct the protocols on the absolute,
and then multiply the sign to obtain the output of signed integer.

\subsection{Satisfying Active Security with Abort}
There are known compilers that convert a passively secure protocol
to an actively secure one (with abort).
The compiler \cite{C:CGHIKL18} and its extension \cite{ACISP:KAHIIMSS19} can be applied to Binary/arithmetic circuit computation,
and each step of our proposed protocols except $\Fbitdecomp$, $\Fmodconv$, and $\Fdiv$
is circuit computation over modulus 2 and $p$.
Therefore, we can obtain actively secure versions of our protocols computing elementary functions
by applying that compiler on modulus 2 and $p$ in parallel.

\section{Secure Deep Neural Network}
The main application we consider in this paper for our secure protocols, is the construction of secure deep neural networks.
We will first give a brief overview of the functions required to implement deep neural networks, and then present secure protocols for these. 
In particular, we propose efficient secure protocols for the softmax activation function and the Adam optimization algorithm for training.

\subsection{Neural Network}
In this paper, we deal with feedforward and convolutional neural networks. 
A network with two or more hidden layers is called a deep neural network, 
and learning in such a network is called deep learning.


A layer contains neurons and the strength of the coupling of neurons between adjacent layers described by parameters $w_i$. 
Learning is a process that (iteratively) updates the parameters to obtain the appropriate output.
\subsubsection{Layer} 
There are several types of layers.
The fully connected layer is computing 
the inner product of the input vector with the parameters.
The convolutional layer is a fully-connected layer 
with removing some parameters and computations.
The max-pooling computes the max value in an input vector
to obtain a representative.
The batch normalization performs normalization and an affine transformation of the input. To normalize a vector $\vec{x}=(x_1,\ldots, x_n)$, we compute 
\begin{equation}
x_{i} \gets \frac{x_i-\mu}{\sqrt{\sigma^2+\epsilon}}, \label{eq:normalization}
\end{equation}
where $\mu$ and $\sigma$ are mean and variance of $\vec{x}$, and $\epsilon$ is a small constant.

\subsubsection{Activation Function}
In a neural network, the activation functions of the hidden layer and 
the output layer are selected according to purpose.

\texthead{ReLU Function}
A popular activation function at the middle layer is the ReLU function defined as $
\textbf{ReLU}(u) = \max(0, u).
$
the function $ \textbf{ReLU}'(u)$, outputting 0 if $u \le 0$ and 1 otherwise, is used as (a substitute of) differentiated ReLU function. 


\texthead{Softmax Function}
Classifications for image identification commonly use the softmax function $\textbf{softmax}(u_i)$ at the output layer. 
The softmax function for classification into $k$ classes is as follows:
\begin{equation}\small
\textbf{softmax}(u_i) = \frac{e^{u_i}}{\sum_{j = 0}^{k-1}e^{u_j}} = \frac{1}{\sum_{j = 0}^{k-1}e^{u_j-u_i}}. \label{eq:softmax}
\end{equation}

\subsubsection{Optimization}

A basic method of parameter update is stochastic gradient descent (SGD).
This method is relatively easy to implement but 
has drawbacks such as slow convergence and the potential for becoming stuck at local maxima.
To address these drawbacks, optimized algorithm have been introduced.
\cite{DBLP:journals/corr/Ruder16} analyzed eight representative algorithms,
and Adam~\cite{DBLP:journals/corr/KingmaB14} was found to be providing particularly good performance.
In fact, Adam is now used in several machine learning framework~\cite{keras, scikitlearn}.


The process of Adam includes the equation
\begin{equation}\small
 W_{t+1} = W_t - \frac{\eta}{\sqrt{\hat{V}_{t+1}}+\epsilon}\circ \hat{M}_{t+1}, \label{eq:adamlasteq}
\end{equation}
where $t$ indicates the iteration number of the learning process,
$W, \hat{V}$, and $\hat{M}$ are matrices, $\circ$ denotes the element-wise multiplication, and $\eta$ and $\epsilon$ are parameters.

\subsection{Secure Protocols for Deep Neural Networks}\label{sec:secureprotNN}
\newcommand{\ashift}{{\mathsf{ashift}}}
\newcommand{\Adam}{{\mathsf{Adam}}}
\newcommand{\Extsign}{{\mathsf{ExtSign}}}
\renewcommand{\ReLU}{{\mathsf{ReLU}}}
\renewcommand{\softmax}{{\mathsf{softmax}}}

The softmax function, Adam, and batch-normalization are quite common and popular algorithms for deep neural network due to their superior performance compared to alternatives.
However, efficient secure protocols for these have been elusive
due to intractability of computing the elementary functions the depend on. 
The softmax function requires exponentiation and inversion, as shown in Eq.~(\ref{eq:softmax}),
and Adam and batch-normalization require the inverse of square roots, as shown in Eq.~(\ref{eq:adamlasteq}) and (\ref{eq:normalization}).
Therefore, the softmax function has often been approximated by a different function \cite{DBLP:conf/sp/MohasselZ17}, which always reduces accuracy, sometimes significantly \cite{keller2020effectiveness}, and only SGD, an elemental optimization method is used.
Although FALCON realized the secure batch-normalization \cite{Falcon}, it is not perfectly secure because it leaks the magnitude of $b = {\sigma^2+\epsilon}$, i.e., $\alpha$ such that $2^\alpha \le b < 2^{\alpha+1}$, to compute $\frac{1}{\sqrt{b}}$.

However, the efficient secure protocols for the elementary functions including exponentiation, division, inversion, and the inverse of square root presented in Section~\ref{section:elementary} allow us to implement secure deep neural network using softmax, Adam, and batch-normalization, as opposed to resorting to approximations and less efficient learning algorithms.


We further prepare building blocks other than the softmax, Adam, and batch-normalization as follows. 
A popular process in layers is matrix multiplication. We apply \cite{C:CGHIKL18} to compute the inner products with the same communication cost of a single multiplication. 
Since the ReLU$^\prime$ function extracts the sign of the input,
we can use the same approach as in Protocol~\ref{alg:extsignabs}.
The ReLU function can be obtained by simply multiplying the input with the output of ReLU$^\prime$.
Secure max-pooling computes the maximum value and a flag (that is required in backpropagation) to indicate which is the maximum value by repeatedly applying the comparison protocol \cite{ACISP:KIMHC18}.

These building blocks are combined to form a secure deep learning system. More details can be found in Appendix~\ref{sec:defernn}, which includes discussion about other ML related techniques.

\section{Experimental Evaluation}\label{sec:experiment}
\smallstartit{Environment.}
%
\begin{table}[t]\footnotesize
\caption{Environment \label{tab:environ}}
\centering
\scalebox{1}{ 
    \begin{tabular}{cl}
    \toprule
    OS& CentOS Linux release 7.3.1611\\
    \hline    
    CPU &  Intel Xeon Gold 6144k (3.50GHz 8 core/ 16 thread) $\times$ 2\\
    \hline    
    Memory& 768 GB\\
    \hline    
     NW &Intel {\if0 Ethernet Controller \fi} X710/X557-AT 10G Ring configuration\\
    \bottomrule   
     \end{tabular}
     }
\end{table}
We implemented our protocols using $p=2^{61}-1$, and instantiated $\Fbitcomp$, $\Fmodconv$, and $\Fqt$ with
the bit-composition, modulus-conversion, and quotient transfer protocols from \cite{ACISP:KIMHC18}, respectively.
We set the statistical security parameter for active-with-abort security to $\kappa=8$.\footnote{While this parameter is relatively small compared to a somewhat more standard parameter like $\kappa=40$ \cite{DBLP:conf/sp/ArakiBFLLNOWW17}, an active adversary will have only a single chance to cheat for the implemented techniques \cite{C:CGHIKL18,ACISP:KAHIIMSS19} 
and an honest party can detect it with probability over $99\%$.}
All experiments are run in the execution environment shown in Table~\ref{tab:environ}, artificially limiting the network speed to 320Mbps and latency to 40ms when simulating a WAN setting.

\subsection{Accuracy and Throughput}
We measured the accuracy and throughput of our division protocol and elementary functions. Due to space limitation, the details of our experiments are deferred to Appendix~\ref{sec:expdivelementary}. In the following, we highlight our key findings.
\begin{itemize}\setlength{\leftskip}{-2mm}
    \item Output of our division protocol is close to real-valued division, with an L1-norm error of $0.335 / 2^t$ for input with offset $t$.
    \item All elementary functions have at least 23-bit accuracy.
    \item The throughput for the elementary functions are an order of magnitude faster than Sharemind~\cite{Ran17} when processing 1M records.\footnote{Note that \cite{Ran17} seems like the most relevant comparison; ABY3 does not implement similar elementary functions, but replaces these with MPC-friendly functions.}
\end{itemize}

\subsection{Secure Training of DNNs}
We measured the performance of training the DNN architectures highlighted below. 
The parameters for Adam in all our experiments are $\beta_1 = 0.09, \beta_2 = 0.999, \eta = 2^{-10} (\eta^\prime = 10), \epsilon = 0$,
which are the recommended parameters in \cite{DBLP:journals/corr/KingmaB14} (except $\epsilon$). 

\smallstart{Network Architectures}
We consider three networks in our experiments: (1) 3DNN, a simple 3-layer fully-connected network introduced in SecureML~\cite{DBLP:conf/sp/MohasselZ17} and used as a benchmark for privacy preserving ML, (2) AlexNet, the famous winner of the
2012 ImageNet ILSVRC-2012 competition \cite{AlexNet} and a network with more than 60 million parameters, and (3) VGG16, the runner-up of the  ILSVRC-2014 competition \cite{VGG16} and a network with more than 138 million parameters.
While the first network is typically used as a performance benchmark for privacy preserving ML, measurements with the latter two networks provide insight into the performance when using larger more realistic networks.


\smallstart{Datasets}
We use two datasets for our experiments: (1) MNIST~\cite{lecun2010mnist}, a collection of 28 x 28 pixel images of hand-written digits typically used for benchmarks, and (2) CIFAR-10~\cite{cifar10}, a collection of colored 32 x 32 pixel images picturing dogs, cats, etc. 
We used MNIST in combination with 3DNN for benchmarking, and CIFAR-10 in combination with the larger networks AlexNet and VGG16.

\smallstart{Comparison}
For experimental comparison with related work, we will focus on the state-of-the-art three-party protocol, FALCON~\cite{Falcon}. We note that the two-party protocols, SecureML~\cite{DBLP:conf/sp/MohasselZ17} and Quotient~\cite{CCS:ASKG19}, are outperformed by any of the three-party protocols by almost an order of magnitude in terms of running time, and among the three-party protocols, FALCON improves upon ABY3, which again is an improvement upon SecureNN. Furthermore, FALCON is the only other related work considering larger networks, AlexNet and VGG16.

Concretely, for all experiments, we ran the code from \cite{Falcon} in the same experimental environment and measured the execution time.
We note that FALCON provides only 13-bit accuracy and sacrifices perfect security for performance, whereas our protocols provide 23-bit accuracy and perfect security.
Furthermore, the code from \cite{Falcon} implements the online phase only, and hence, the measurements do not include the corresponding offline phase, which would make a significant contribution to the total running time. 
Lastly, the code does not update the parameters of the model, which makes the accuracy of the obtained model unclear. 
We emphasize that the measurements for our protocols are for the \textit{total} running time and a fully trained model.
Despite this, we treat the obtained measurements as comparable to ours.
This is in favor of FALCON.

\smallstart{Results for 3DNN}
\begin{table}[t]\footnotesize
\caption{Comparison of training time of 3DNN over the MNIST dataset.}
\vspace{-1mm}
    \centering
    \begin{tabular}{c wc{4em} c wc{2em} r r}
\toprule
&
Security/NW   
  & Methods & Epochs & 
          Time [s]
        &
        Accuracy
        {\scriptsize {[}\%{]}}
         \\
\hline  
FALCON  &  
\multirow{2}{*}{
\begin{tabular}{l}
Passive/LAN
\end{tabular}
}
 & SGD & $15$ & $780$ & - \\ 
\textbf{Ours} & & Adam    & $1$ & $117$ & $95.64$ \\
\hline 
FALCON &  \multirow{2}{*}{
Active/LAN
}
 & SGD & $15$ & $2,355$ & -\\ 
\textbf{Ours} && Adam    & $1$ & $570$ & $95.61$ \\
\hline
FALCON &  \multirow{2}{*}{
\begin{tabular}{l}
Passive/WAN
\end{tabular}
}
& SGD & $15$ & $16,110$ & - \\
\textbf{Ours} && Adam    & $1$ & $4,537$ & $95.64$ \\
\hline 
FALCON & 
\multirow{2}{*}{
\begin{tabular}{l}
Active/WAN
\end{tabular}
}
&SGD & $15$ & $37,185$ & -\\ 
\textbf{Ours} && Adam    & $1$ & $11,516$ & $95.61$ \\
\bottomrule
    \end{tabular}
    \label{tab:compareall}
\par
\end{table}
We measured the running time and accuracy for training 3DNN on the MNIST dataset for passive and active security, in the LAN and WAN settings. 
The results for our protocols and FALCON can be seen in Table~\ref{tab:compareall}. 
Compared to FALCON, ours is between $3.2$ to $6.7$ times faster, depending on the setting. We again highlight that these results are achieved despite the advantages provided to FALCON in this comparison (measuring online time only, 13-bit vs. 23-bit accuracy, and imperfect security). 
For reference, we note that ours is only a factor of less than $6$ slower than training \textit{in the clear} using MLPClassifier from~\cite{scikit-learn} ($17.7$ seconds, $95.54$ \% accuracy) on a single machine.

\smallstart{Results for AlexNet and VGG16}
In the original paper of FALCON~\cite{Falcon}, the total running time for training on AlexNet and VGG16 was estimated through extrapolation since these networks require a significant amount of computation for training, even in the clear. We follow this method to estimate the running time of ours and FALCON (re-evaluated in our environment) in the same way.

In Table \ref{tab:compare-epochs}, we show the measured running time to complete a single epoch for AlexNet and VGG16 using the CIFAR-10 dataset, both for passive and active security, as well as in the LAN and WAN settings. 
The table include measurements for both FALCON and our protocols.
Note, however, that the time to complete a single epoch is not indicative of the overall performance difference between FALCON and our framework, as the underlying optimization methods are different, and require a different number of epochs to train a network achieving a certain prediction accuracy. 


\begin{table}[t]\footnotesize
\caption{Measured running time per epoch for training AlexNet and VGG16 on CIFAR-10.}
\vspace{-1mm}
    \centering
    \begin{tabular}{cccrr}
\toprule
        & Security & Setting & AlexNet [s] & VGG16 [s]  \\
\hline  
FALCON & Passive & LAN &  $10,892$ & $523,127$ \\
\textbf{Ours}& Passive & LAN & $3,139$ & $43,150$ \\
\hline
FALCON & Active & LAN &  $41,537$ & $2,051,751$ \\
\textbf{Ours}& Active & LAN & $15,021$ & $161,481$ \\
\hline  
FALCON & Passive & WAN &  $23,489$ & $575,699$ \\
\textbf{Ours}& Passive & WAN & $49,833$ & $347,928$ \\
\hline  
FALCON & Active & WAN &  $75,838$ & $2,240,515$ \\
\textbf{Ours}& Active & WAN & $159,781$ & $1,293,226$ \\
\bottomrule
    \end{tabular}
    \label{tab:compare-epochs}
\end{table}

\begin{figure}[t]
\centering
     \begin{subfigure}[b]{\columnwidth}
         \centering
         \includegraphics[scale=0.45]{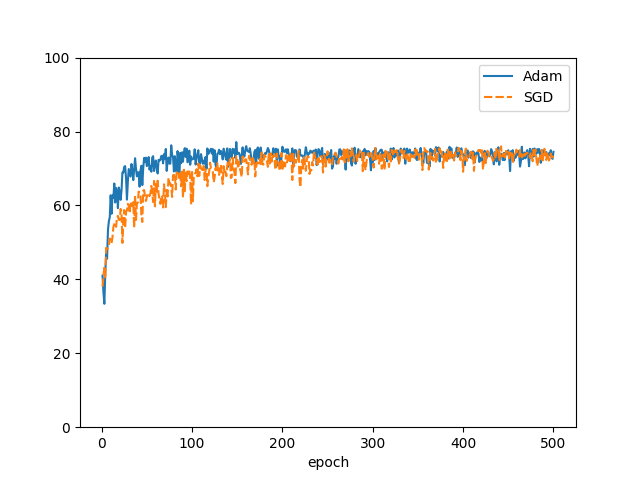}
         \caption{AlexNet}
         \label{fig:alexcifar}
     \end{subfigure}
     \vspace{-4mm}
     \begin{subfigure}[b]{\columnwidth}
         \centering
         \includegraphics[scale=0.45]{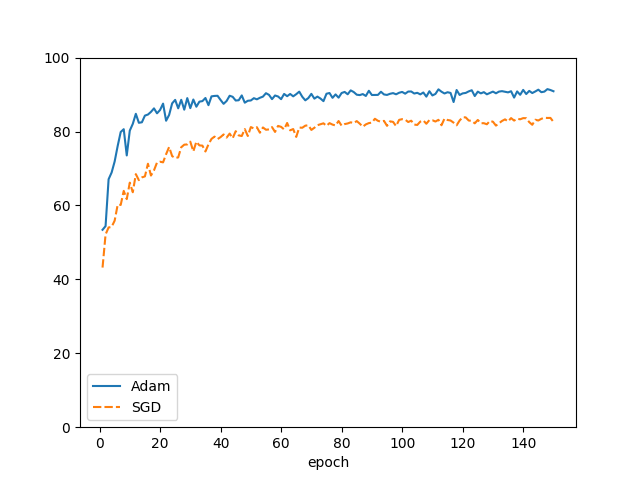}
         \caption{VGG16}
         \label{fig:vggcifar}
     \end{subfigure}
        \caption{Accuracy of AlexNet and VGG16 trained with Adam and SGD on CIFAR-10.}
        \label{fig:plaintext-accuracy}
\end{figure}

To determine the number of epochs needed for Adam (implemented in our framework) and SGD (implemented in FALCON), we ran Adam and SGD for AlexNet and VGG16 on CIFAR-10 in the clear, and measured the achieved accuracy. The results are illustrated in Figure \ref{fig:plaintext-accuracy}. For AlexNet, we see that accuracy converges towards $75\%\sim 78\%$, with Adam achieving a maximum of 77.15\% and SGD a maximum of 75.98\% in our test. We note that Adam achieves an accuracy exceeding 70\% after 25 epochs, whereas SGD requires 107 epochs. For VGG16, we see that Adam significantly outperforms SGD, and after relatively few epochs, achieves an accuracy not obtained by SGD, even after 150 epochs. We note that achieving an accuracy exceeding 75\% requires 6 and 23 epochs for Adam and SGD, respectively, whereas an accuracy of 80\% requires 8 and 45 epochs, respectively.

Based on the observations above, we estimate the running time of achieving an accuracy of 70\% for AlexNet and 75\% for VGG16, for active and passive security in the LAN and WAN setting. The result is shown in Table \ref{tab:compare-total}. We see that in the LAN setting, the total running time of our framework outperforms the online phase of FALCON with a factor of about $12\sim14$ for AlexNet and $46\sim48$ for VGG16, whereas in the WAN setting, the factors are about 2 and 6, respectively. 

\begin{table}[t]\footnotesize
\caption{Estimated running time for training of AlexNet (70\% accuracy) and VGG16 (75\% accuracy) on CIFAR-10.}
\vspace{-1mm}
    \centering
    \begin{tabular}{cccrr}
\toprule
        & Security & Setting & AlexNet [h] & VGG16 [h]  \\
\hline  
FALCON & Passive & LAN &  $324$ & $3,342$ \\
\textbf{Ours}& Passive & LAN & $22$ & $72$ \\
\hline
FALCON & Active & LAN &  $1,235$ & $13,108$ \\
\textbf{Ours}& Active & LAN & $104$ & $269$ \\
\hline  
FALCON & Passive & WAN &  $698$ & $3,678$ \\
\textbf{Ours}& Passive & WAN & $346$ & $580$ \\
\hline  
FALCON & Active & WAN &  $2,254$ & $14,314$ \\
\textbf{Ours}& Active & WAN & $1,110$ & $2,155$ \\
\bottomrule
    \end{tabular}
    \label{tab:compare-total}
\end{table}

The above comparison illustrates the advantage of the approach taken in our framework; by constructing efficient (and highly accurate) protocols that allow advanced ML algorithms such as Adam to be evaluated, despite these containing ``MPC-unfriendly functions'', we gain a significant advantage in terms of overall performance compared to previous works like FALCON, that attempt to achieve efficiency by simplifying the underlying ML algorithms, and optimizing the evaluation of these. 
As shown, the advantage when considering larger more realistic networks can in some cases be significantly more pronounced than suggested by the evaluation results on benchmark networks such as 3DNN, which is illustrated by the obtained $46$ times faster evaluation of VGG16 in the LAN setting.
We again highlight that this is obtained despite the advantages offered to FALCON in the comparison.

\smallstart{Note on Trident}
Finally, for completeness, we note that Trident~\cite{NDSS:RachuriS20} improves upon the online phase of ABY3 by increasing the number of servers to four and pushing more of the computation to the offline phase.
Note that being a four-party protocol (tolerating a single corruption), Trident obviously increases cost in terms of the required number of servers, but also weakens security compared to the above mentioned three-party protocols, and is hence not directly comparable to our framework.
Nevertheless, from the measurements which are provided in \cite{NDSS:RachuriS20}, we estimate that, for active security, the \textit{online} phase of Trident is somewhat faster in the LAN setting and somewhat worse in the WAN setting compared to the \textit{total} running time for our protocols when considering the simple 3DNN network.\footnote{From the measurements reported in~\cite{NDSS:RachuriS20}, we estimate that the online phase of Trident in their environment would require 306s and 30264s to train 3DNN on the MNIST dataset in the LAN and WAN setting, respectively.
The corresponding total time for our protocols are 570s and 11516s, respectively.}
However, including the offline phase, which is significant for Trident, will add considerably to the running time\footnote{Note that the offline phase in Trident is slower than the semi-honest ABY3 implementation (see \cite{NDSS:RachuriS20}[Appendix E.B]), which is already heavy. 
Furthermore note that the offline communication cost in Trident is the same or larger than in the online phase for all the 12 protocols in \cite{NDSS:RachuriS20}, except bit extraction.}.
This strongly indicates that, despite being a four-party protocol, Trident offers worse overall performance than our protocols, in particular in the WAN setting, while at the same time relying on simplifications of the underlying ML learning algorithms.
Additionally, Trident does not implement batch normalization required for AlexNet, and does not report any measurements for larger networks, for which we expect our framework to have a greater advantage. Since the source code is furthermore not available, such measurements are not easily obtainable.

\section{Conclusion}
In this paper, we proposed a framework that enables efficient and secure evaluation of ML algorithms via three-party protocols for MPC-unfriendly computations.
We first proposed a new division protocol, which enables efficient and accurate fixed-point arithmetic computation,
and based on this, efficient protocols for machine learning, such as inversion, square root extraction, and exponential function evaluation.
These protocols enable us to efficiently compute modern ML algorithms such as Adam and the softmax function as is. 
As a result, we obtain secure DNN training that outperforms state-of-the-art three-party systems in all tested settings, with the most pronounced advantage for large networks in the LAN setting.

\bibliographystyle{IEEEtranS}
\bibliography{abbrev3,crypto,others,crypt}

\appendices

\section{Deferred Details on Share Conversions}
\label{sec:shareconversion}
This section provides details for share conversions, deferred from Section~\ref{sect:ss}.
Our protocols will utilize the following share conversions.

\subsubsection*{$\pash{a} \to \tash{a}$}
Let $\pash{a}_i=(a_i,a_{i+1})$ be $P_i$'s share for $i=1,2,3$.
The conversion from $\pash{a}$ to $\tash{a}$, which we denote $\tash{a} \gets \ptotconvert(\pash{a})$, is a local operation:
$P_1$ and $P_2$ set $\tash{a}_1 := a_1$ and $\tash{a}_2 := a_2 + a_3$, respectively.

\subsubsection*{$\tash{a} \to \pash{a}$}
The conversion from $\tash{a}$ to $\pash{a}$, which we denote $\pash{a} \gets \ttopconvert(\tash{a})$, is achieved via a simple protocol from \cite{ACISP:KIMHC18}:
$P_1$ and $P_2$ secret-share their shares $\tash{a}_1$ and $\tash{a}_2$ using a $\pash{\cdot}$-sharing, and send $\pash{\tash{a}_1}_i$ and $\pash{\tash{a}_1}_i$, respectively, to party $P_i$.
Each $P_i$ adds the received shares as $\pash{a}_i = \pash{\tash{a}_1}_i + \pash{\tash{a}_2}_i$ locally.
We can further optimize this algorithm by using a pseudo-random number generator \cite{DBLP:journals/iacr/ChidaHIKKP19}, to achieve a communication cost of two field elements among three parties in a single round. 

\subsubsection*{$\pash{a} \to \bash{a_\ell},\ldots,\bash{a_1}$}
This conversion decomposes 
a shared secret $a$ into shares of its bit representation $a_\ell,\ldots,a_1$, which is also known as bit-decomposition.
Note that we only decompose the least significant $\ell$ bits of $a$.
The ideal functionality for bit-decomposition, which we denote  $\Fbitdecomp$, is given in
Functionality~\ref{func:bitdecomp}.
A protocol that securely computes $\Fbitdecomp$ in the presence of passive and active adversaries appear in \cite{ACISP:KIMHC18}.

\begin{figure}[htbp]
\centering
\framebox[\width][c]{
    \small\,
    \hbox{
    \begin{varwidth}[c]{0.43\textwidth}
    \begin{myfunctionality}
    [$\Fbitdecomp$ -- Bit decomposition]
    \label{func:bitdecomp}
    \end{myfunctionality}
    {Upon receiving $\pash{a}$, $\Fbitdecomp$ reconstructs $a$, generates shares $(\bash{a_1},\ldots, \bash{a_\ell})$, 
where $a = \sum_{i=1}^{\ell}2^{i-1}a_i$ and sends $(\bash{a_1}_i,\ldots, \bash{a_\ell}_i)$ to $P_i$.}
    \end{varwidth}
    \,
    }
}
\end{figure}

\subsubsection*{$\bash{a_\ell},\ldots,\bash{a_1} \to \pash{a}$}
This conversion constructs shares of a secret $a$ from shares of its bit representation $a_\ell,\ldots,a_1$, which is also known as bit-composition.
The ideal functionality for bit-composition, which we denote $\Fbitcomp$, is given in Functionality~\ref{func:bitcomp}.
In Appendix~\ref{sec:bitcomp}, we present an efficient bit-composition protocol based on quotient transfer presented below.
This is a modified version for modulus $p$ while the known bit composition protocol \cite{CCS:ABFKLO18} works in only modulus $2^n$.
\begin{figure}[htbp]
\centering
\framebox[\width][c]{
    \small\,
    \hbox{
    \begin{varwidth}[c]{0.43\textwidth}
    \begin{myfunctionality}
    [$\Fbitcomp$ -- Bit composition]
    \label{func:bitcomp}
    \end{myfunctionality}
    {Upon receiving $\bash{a_1},\ldots,\bash{a_\ell}$, $\Fbitdecomp$ reconstructs $a_1,\ldots,a_\ell$, computes $a:= \sum_{i=1}^\ell a_i \cdot 2^i$, generates shares $\pash{a}$, and sends $\pash{a}_i$ to $P_i$.}
    \end{varwidth}
    \,
    }
}
\end{figure}

\subsubsection*{$\bash{a} \to \pash{a}$}
This conversion essentially changes the modulus of the underlying field of the shares while maintaining the secret i.e.
 shares $\bash{a}_i \in \Zset_2$ for a secret $a \in \bit$ are converted to shares $\pash{a}_i \in \Zset_p$.  
The ideal functionality of this modulus-conversion is given in
Functionality~\ref{func:modconv}.
Protocols that securely compute $\Fmodconv$ in the presence of passive and active adversaries appear in \cite{ACISP:KIMHC18}.

\begin{figure}[htbp]
\centering
\framebox[\width][c]{
    \small\;
    \hbox{
    \begin{varwidth}[c]{0.43\textwidth}
    \begin{myfunctionality}
    [$\Fmodconv$ -- Modulus conversion]
    \label{func:modconv}
    \end{myfunctionality}
    {Upon receiving $\bash{a}$, $\Fmodconv$ reconstructs $a$, generates shares $\pash{a}$, and sends $\pash{a}_i$ to $P_i$.}
    \end{varwidth}\;
    }
}
\end{figure}

\section{Quotient Transfer Protocol}\label{sect:QTprotocol}

We describe the quotient transfer protocols for $\pash{\cdot}$, $\tash{\cdot}$, and $\bash{\cdot}$.

Informally speaking, the main idea of this protocol is that 
if we use an odd prime and the secret's LSB is zero,
the addition of the truncated shares' LSBs corresponds to $q$.

The quotient transfer protocol uses multiplication and modulus conversion protocols.
In the protocol, we describe them as functionalities $\Fmult$ and $\Fmodconv$.
Please see \cite{ACISP:KIMHC18} for their instantiations.

In the presence of passive adversaries,
we use $\tash{a}$ as an input of the quotient transfer protocol.
Because $\tash{a}$ consists of two sub-shares, the quotient $q$ is 0 or 1
and the (single) LSB must be 0.

\begin{algorithm}[t]\small 
\caption{Quotient Transfer for $\tash{\cdot}$}
 \label{alg:qt}
 \begin{algorithmic}[1]
  \Functionality $\tash{q} \gets \QT(\tash{a})$
    \Require $\tash{a}$ where $a$ is a multiple of $2$.
    \Ensure $\tash{q}$ where $\tash{a}_1 + \tash{a}_2 = a + qp$
    \State $P_0$ and $P_1$ secret-share LSBs of $\tash{a}_1$ and $\tash{a}_2$ in modulo 2, respectively. 
        Let them be $\bash{\tash{a}_1^{(1)}}$ and $\bash{\tash{a}_2^{(1)}}$.
    \State $\bash{q} := \bash{\tash{a}_1^{(1)}} \oplus \bash{\tash{a}_2^{(1)}}$.
    \State $\pash{q} \gets \Fmodconv(\bash{q})$
    \State $\tash{q} \gets \ptotconvert(\pash{q})$
    \State Output $\tash{q}$ 
 \end{algorithmic}
\end{algorithm}

In the presence of active adversaries,
we use $\pash{a}$ as an input of the quotient transfer protocol.
Because $\pash{a}$ consists of three sub-shares,
the quotient $q$ is 0, 1, or 2, and the second LSBs must be $0$s to contain $2$.
The step 3 and the last term of step 4 come from the fact that
the carry of $a_1, a_2, a_3$ is $(a_1\oplus a_3)(a_2 \oplus a_3) \oplus a_3$.
This protocol is secure against an active adversary with abort 
using general compiler such as \cite{C:CGHIKL18}.
Note, in the step 1, the ``share of sub-shares'' can be generated locally. 
For details, see Section 4.4 in \cite{ACISP:KIMHC18}.

\begin{algorithm}[t]\small 
\caption{Quotient Transfer for $\pash{\cdot}$}
 \label{alg:qt2}
 \begin{algorithmic}[1]
  \Functionality $\pash{q} \gets \QT(\pash{a})$
    \Require $\pash{a}$ where $a$ is a multiple of $4$.
    \Ensure $\pash{q}$ where $a_1 + a_2 + a_3 = a + qp$
    \State $P_0$ and $P_1$ \emph{locally} generate shares of the second LSBs of $a_1$, $a_2$, and $a_3$ in modulo 2, respectively.
        Let them be $\bash{a_1^{(1)}}$, $\bash{a_1^{(2)}}$, $\bash{a_2^{(1)}}$, $\bash{a_2^{(2)}}$, $\bash{a_3^{(1)}}$, and $\bash{a_3^{(2)}}$.
    \State $\bash{q_1}:= \bash{a_1^{(1)}} \oplus \bash{a_2^{(1)}} \oplus \bash{a_3^{(1)}}$.
    \State  $\bash{c} \gets (\bash{a_1^{(1)}}\oplus\bash{a_3^{(1)}}) \cdot (\bash{a_2^{(1)}}\oplus\bash{a_3^{(1)}}) \oplus \bash{a_3^{(1)}}$.
    \State $\bash{q_1} := \bash{a_1^{(2)}} \oplus \bash{a_2^{(2)}} \oplus \bash{a_3^{(2)}} \oplus \bash{c}$.
    \State $\pash{q_1} \gets \Fmodconv(\bash{q_1})$
    \State $\pash{q_2} \gets \Fmodconv(\bash{q_2})$
    \State Output $\pash{q}:= \pash{q_1} + 2\pash{q_2}$ 
 \end{algorithmic}
\end{algorithm}

We can also introduce the quotient transfer protocol for $\bash{\cdot}$ \cite{CCS:ABFKLO18}.
This protocol computes the carry, which means whether or not there are at least two sub-shares are $1$.

\begin{algorithm}[t]\small 
\caption{Quotient Transfer for $\bash{\cdot}$}
 \label{alg:qt3}
 \begin{algorithmic}[1]
  \Functionality $\bash{q} \gets \QT(\bash{a})$
    \Require $\bash{a}$ (no restriction of $a$)
    \Ensure $\bash{q}$ where $a_1 + a_2 + a_3 = a + 2q$
    \State $P_0$ and $P_1$ \emph{locally} generate shares of $a_1$, $a_2$, and $a_3$ in modulo 2, respectively.
        Let them be $\bash{a_1}$, $\bash{a_2}$, and $\bash{a_3}$.
    \State $\bash{q} \gets (\bash{a_1} \oplus \bash{a_3}) \cdot (\bash{a_2}\oplus \bash{a_3}) \oplus \bash{a_3}$.
    \State Output $\bash{q}$ 
 \end{algorithmic}
\end{algorithm}

\section{Functionalities of Division Protocols}\label{sect:functionality}
We give the functionalities for division by a public value,
$\Fdiv$ and $\Fdivgen$, corresponding to division in the specific and general case, respectively, in Functionality~\ref{func:div-specific} and Functionality~\ref{func:div}.

\functionality
{$\Fdiv$ -- Division by a public value}
{func:div-specific}
{htbp}
{Upon receiving $\tash{a}_1$ and $\tash{a}_2$,
$\Fdiv$ reconstructs $a \equiv \tash{a}_1 + \tash{a}_2$ and computes 
$r_a = a \mod d$, and $r_1 = \tash{a}_1 \mod d$.
Then, $\Fdiv$ sets $b$ as follows:
\begin{itemize} \setlength{\leftskip}{-2mm}
    \item $b = a/d$ if $\Big((\tash{a}_1 \le a)\wedge (r_a < r_1)\Big) 
        \vee \Big((a < \tash{a}_1)\wedge (r_a-1 < r_1)\Big)$, \\ 
    \item $b = a/d + 1$ otherwise.
\end{itemize}
Then, $\Fdiv$ randomly picks $\tash{b}_1 \gets F_p$, sets $\tash{b}_2 \equiv b - \tash{b}_1$, and hands the parties $P_1$ and $P_2$ their shares $\tash{b}_1$ and $\tash{b}_2$, respectively.
}

\functionality
{$\Fdivgen$ -- Division by a public value in general case}
{func:div}
{htbp}
{Upon receiving $\tash{a}_1$ and $\tash{a}_2$,
$\Fdiv$ reconstructs $a \equiv \tash{a}_1 + \tash{a}_2$ and computes 
$r_a = a \mod d$, $r_p = p \mod d$, and $r_1 = \tash{a}_1 \mod d$.
Then, $\Fdiv$ sets $b$ as follows:
\begin{itemize} \setlength{\leftskip}{-2mm}
    \item $b = a/d$ if $\Big((\tash{a}_1 \le a) \wedge (r_a < r_1 \le r_p)\Big) \vee \Big((a < \tash{a}_1)\wedge \big((r_a+r_p < r_1) \vee (r_a+r_p-d < r_1 \le r_p)\big)\Big)$. 
    \item $b = a/d + 1$ if $\Big( (\tash{a}_1 \le a) \wedge \big((r_1 \le r_a)\wedge(r_1 \le r_p)\big) \vee \big((r_a < r_1)\wedge(r_p < r_1)\big)\Big) \vee
         \Big( (a < \tash{a}_1) \wedge \big( (r_p< r_1 \le r_a + r_p) \vee (r_1 \le r_a+r_p-d) \big) \Big)$.
    \item $b = a/d + 2$ if $(\tash{a}_1 \le a) \wedge (r_p < r_1 \le r_a)$.
\end{itemize}
Then, $\Fdiv$ randomly picks $\tash{b}_1 \gets F_p$, sets $\tash{b}_2 \equiv b - \tash{b}_1$, and hands the parties $P_1$ and $P_2$ their shares $\tash{b}_1$ and $\tash{b}_2$, respectively.}

\section{Division Protocol for Signed Integers}\label{sect:divisionsigned}
We extend our division protocols to signed integers in Protocol~\ref{alg:divisionLSSsigned}.

\begin{algorithm}[htbp] 
\caption{Secure Division by Public Value in $\pash{\cdot}$ with \emph{Signed} Integers}
 \label{alg:divisionLSSsigned}
 \begin{algorithmic}[1]
  \Functionality $\pash{c}\gets \Div^\textsf{S}_{(2,3)}(\pash{a}, d)$
   \Require Share of dividend $\pash{a}$ and (public) divisor $d$, 
    where $-2^{\left|p\right|-2}-r_\omega \le  a \le 2^{\left|p\right|-2}-r_\omega-1$
  \Ensure $\pash{c}$, where $c \approx \frac{a}{d}$
  \Parameter $\omega := \ceil{\frac{2^{\left|p\right|-2}}{d}}$ and $r_\omega$ such that $\omega d = 2^{\left|p\right|-2} + r_\omega$
  \State $\pash{b} \gets \Div_{(2,3)}(\pash{wd+a}, d)$ 
  \State $\pash{c} \gets \pash{b} - \pash{w}$ 
  \State Output $\pash{c}$
 \end{algorithmic}
\end{algorithm}

\section{Precise Analysis of Our Division Protocol}\label{sect:divgeneral}

\subsection{Specific Case}
First, we focus on 
an the important case (for our application) in which $p$ is a Mersenne prime and $d$ is a power of $2$.

In that case, $r_p = d-1$ and Eq.~(\ref{eq:division}) is
\begin{align}\label{eq:division-specific}
    & \alpha_a  -q + 1 + \frac{r_a + q(d-1) - r_1 -r_2}{d} + r_1/d  \nonumber \\
    &= \alpha_a  -q + 1 + \frac{r_a + q(d-1) - r_1 -r_2}{d}
\end{align}
since $r_1 <d$.

Next, we separate the cases of $q=0$ and $q=1$.
If $q=0$ (that means $\tash{a}_1 \le a$), 
Eq.~(\ref{eq:division-specific}) is 
\[
    \alpha_a + 1 + \frac{r_a - r_1 -r_2}{d}.
\]
Since $\tash{a}_1 + \tash{a}_2 \equiv a$ and $q=0$, 
the equation $(\alpha_1 d + r_1) + (\alpha_2 d + r_2) = \alpha_a d + r$ holds 
and converts to
\[
\frac{r_a-r_1-r_2}{d} = \alpha_1 + \alpha_2 - \alpha_a.
\]
Since $\alpha_1$, $\alpha_2$, and $\alpha_a$ are integers, $r - r_1 -r_2$ must be a multiple of $d$. 
In addition, since $r$, $r_1$, and $r_2$ are less than $d$,
$r_a-r_1-r_2$ is either 0 or $-d$. Precisely, 
\begin{align*}
\frac{r_a - r_1 - r_2}{d} = 
\begin{cases}
    -1 &\text{if } r_a < r_1 \\
    0 &\text{otherwise}.
\end{cases}
\end{align*}
Therefore, in the case of $q=0$, the output is as follows.
\begin{align}\label{eq:q0-specific}
 \alpha_a + 1 + \frac{r_a - r_1 -r_2}{d} 
 = 
\begin{cases}
    \alpha_a  & \text{if } r_a < r_1 \\
    \alpha_a + 1 & \text{otherwise}
\end{cases}.
\end{align}

We then switch to the case of $q=1$ (that means $a < \tash{a}_1$).
Eq.~(\ref{eq:division-specific}) is 
\[
    \alpha_a  + \frac{r_a + r_p - r_1 -r_2}{d}.
\]
Since $\tash{a}_1 + \tash{a}_2 \equiv a$ and $q=1$, 
$(\alpha_1 d + r_1) + (\alpha_2 d + r_2) = (\alpha_a d + r_a) + (\alpha_p d + r_p)$
and 
\[
\frac{r_a + r_p - r_1 - r_2}{d} = \alpha_1 + \alpha_2 - (\alpha_a+\alpha_p).
\]
Since all the terms in the right equation are integers, 
$r_a + r_p - r_1 -r_2$ is a multiple of $d$.
In addition, since $r_a$, $r_1$, and $r_2$ are less than $d$ and $r_p =d-1$,
$r_a + r_p - r_1- r_2$ can be either 0 or 1.
Precisely, 
\begin{align*}
\frac{r_a+r_p - r_1 - r_2}{d} = 
\begin{cases}
    0 &\text{if } r_a - 1 < r_1 \\
    1 &\text{otherwise},
\end{cases}
\end{align*}
and we have the following in the $q=1$ case.
\begin{align}\label{eq:q1-specific}
 \alpha_a  + \frac{r_a + r_p - r_1 -r_2}{d} 
 = 
    \begin{cases}
    \alpha_a  & \text{if } r_a-1 < r_1 \\
    \alpha_a + 1 & \text{otherwise } .
    \end{cases}
\end{align}

Let $b$ be the secret of our protocol's output and recall that $\alpha_a = a/d$. From Eq.~(\ref{eq:q0-specific}) and \ref{eq:q1-specific}, we conclude that
\begin{align}\label{eq:result-specific}
b = 
    \begin{cases}
        a/d  & \text{if } 
            \begin{array}{l}
                \Big((\tash{a}_1 \le a)\wedge (r_a < r_1)\Big)  \\
                \vee \Big((a < \tash{a}_1)\wedge (r_a-1 < r_1)\Big) 
            \end{array}
            \\
        a/d + 1 & \text{otherwise } 
    \end{cases}
\end{align}

\subsection{General Case}
On input an additive share $\tash{a}$ and divisor $d$, 
we show that our division protocol outputs $\tash{a/d}$, $\tash{a/d + 1}$, or $\tash{a/d + 2}$ 
for a general $p$ and $d$.
We first consider the $q=0$ case. In this case, Eq.~(\ref{eq:division}) is 
\[
\alpha_a  + 1 + \frac{r_a - r_1 -r_2}{d} + (r_1-r_p +d-1)/d.
\]
For the term of $\frac{r_a-r_1-r_2}{d}$,
the same discussion in the specific case holds.

We then consider the last term $(r_1-r_p +d-1)/d$. 
If $r_1 -r_p \le 0$, $0 \le r_1-r_p +d-1 < d$ since $-(d-1) \le r_1 -r_p$. 
Otherwise, $d \le  r_1-r_p +d-1 < 2d$ since $r_1 -r_p \le d-1$.
Therefore,
\begin{align}\label{eq:thirdterm}
(r_1 -r_p +d-1)/d = 
\begin{cases}
    0 & \text{if } r_1 \le r_p \\
    1 & \text{otherwise}
\end{cases}
\end{align}
Therefore,
we have the following equation in the $q=0$ case.
{
\begin{align}\label{eq:q0}
& \alpha_a  + 1 + \frac{r_a - r_1 - r_2}{d} + (r_1 +d-1-r_p)/d \nonumber\\
& = 
\begin{cases}
    \alpha_a  & \text{if } r_a < r_1 \le r_p \\
    \alpha_a + 1 & \text{if } 
        \begin{alignedat}{1}
          & \big((r_1 \le r_a)\wedge(r_1 \le r_p)\big)\\
    & \vee \big((r_a < r_1)\wedge(r_p < r_1)\big)\\
    \end{alignedat}\\
    \alpha_a + 2 & \text{if } r_p < r_1 \le r_a
\end{cases}.
\end{align}
}

Next, we consider the $q=1$ case. In this case, Eq.~(\ref{eq:division}) is 
\[
\alpha_a  + \frac{r_a + r_p - r_1 - r_2}{d} + (r_1 -r_p+d-1)/d.
\]
The same discussion as in the specific case holds and $\frac{r_a + r_p - r_1 - r_2}{d}$ must be a multiple of $d$.
However, $r_p$ can be small in the general case, and it affects the possible values of this term;
namely, $r_a + r_p - (r_1+r_2)$ can be $-1$ in addition to 0 and 1.
Precisely, 
{
\begin{align}\label{eq:q1second}
\frac{r_a+r_p - r_1 - r_2}{d} = 
\begin{cases}
    -1 &\text{if } r_a + r_p < r_1 \\
    0 &\text{if } r_a+r_p-d < r_1 \le r_a + r_p \\
    1 &\text{otherwise}
\end{cases}
\end{align}
}
The term $(r_1 -r_p +d-1)/d$ is the same as Eq.~(\ref{eq:thirdterm}).
Therefore, combining the conditions of Eq. (\ref{eq:thirdterm}) and \ref{eq:q1second},
{
\begin{align}\label{eq:q1}
& \alpha_a  + \frac{r_a + r_p - r_1 -r_2}{d} + (r_1-r_p +d-1)/d \nonumber\\
& = 
    \begin{cases}
    \alpha_a  & \text{if } (r_a+r_p < r_1) \vee (r_a+r_p-d < r_1 \le r_p) \\
    \alpha_a + 1 & \text{if } (r_p< r_1 \le r + r_p)
         \vee (r_1 \le r_a+r_p-d)
    \end{cases}.
\end{align}
}

Let $b$ be the secret of our protocol's output and recall that $\alpha_a = a/d$. 
From Eq.~(\ref{eq:q0}) and (\ref{eq:q1}), we conclude that $b = a/d$ if
{
\begin{align*}
        & \Big((\tash{a}_1 \le a) \wedge (r_a < r_1 \le r_p)\Big) \vee\\
        & \Big((a < \tash{a}_1)\wedge \big((r_a+r_p < r_1) \vee (r_a+r_p-d < r_1 \le r_p)\big)\Big),
\end{align*}
}
$b=a/d + 1$ if
{\footnotesize
\begin{align*}
    & \Big( (\tash{a}_1 \le a) \wedge \big((r_1 \le r_a)\wedge(r_1 \le r_p)\big) \vee \big((r_a < r_1)\wedge(r_p < r_1)\big)\Big)\\
    & \vee \Big( (a < \tash{a}_1) \wedge \big( (r_p< r_1 \le r_a + r_p) \vee (r_1 \le r_a+r_p-d) \big) \Big),
\end{align*}
}
and $b=a/d + 2$ otherwise.



\section{Probability of Each Output for Random Shares}\label{sect:divprob}
In this section, we specify how the output of our passively secure division protocol 
depends on $d$, $a$, $p$, $\tash{a}_1$ (and their dependent variables $\alpha_a$, $r_a$, $\alpha_p$, $r_p$, $\tash{a}_2$, $r_1$, and $r_2$).
In this section, we assume that $\tash{a}_1$ is uniformly random in mod $p$. 
This is the case for our application -- multiply-then-truncate.
An output of multiplication protocol is uniformly random share,
and an input of the division protocol is always the output of a multiplication protocol.

\subsection{Specific Case}
As in the correctness discussion, we first focus on 
the case in which $p$ is a Mersenne prime and $d$ is a power of 2.

From Eq.~(\ref{eq:result-specific}),
we count integers that satisfying 
$(\tash{a}_1 \le a)\wedge (r_a < r_1)$ or $(a < \tash{a}_1)\wedge (r_a-1 < r_1)$.
We observe that $(\tash{a}_1 \le a)\wedge (r_a < r_1)$ is true if 
\[
\tash{a}_1 \in \{md + n \mid 0 \le m \le \alpha_a-1, r_a+1 \le n \le d-1\}.
\]
The number of integers satisfying the above is $\alpha_a(d-r_a-1)$.
We then observe that $(a < \tash{a}_1)\wedge (r_a-1 < r_1)$ is true if 
\begin{align*}
\tash{a}_1 \in & \{md + n \mid \alpha_a \le m \le \alpha_p, r_a \le n \le d-1\} \\
    & \setminus \{\alpha_a d + r_a, \alpha_p d + d-1\}.
\end{align*}
The number of integers satisfying the above is $(\alpha_p-\alpha_a + 1)(d-r_a)-2$.

Therefore, the probability that the output is $a/d$ is 
\begin{align*}
    & \frac{\alpha_a(d-r_a-1) + (\alpha_p-\alpha_a + 1)(d-r_a)-2}{p} \\
    &= \frac{\alpha_p d + (d-1) -\alpha_p r_a -\alpha_a + r_a -1}{p}\\
    &= \frac{p - r_a(\alpha_p-1) -\alpha_a -1}{p},
\end{align*}
and the probability that the output is $a/d +1$ is 
\begin{align*}
    & 1- \frac{p -r_a(\alpha_p-1) -\alpha_a -1}{p} 
    = \frac{r_a(\alpha_p-1) + \alpha_a + 1}{p}
\end{align*}





\subsection{General Case}
We show the probability of each output for general $p$ and $d$,
including when $d$ is not a power of $2$.
The probability depends on two relations:
magnitude relations between $r$ and $r_p$, and $d$ and $r_a+r_p$.

\subsection*{Case 1: $r_p<r_a$ and $r_a+r_p<d$}
First, we consider the $q=0$ case, \ie $\tash{a}_1 \le a$. If $r_p < r_a$, Eq.~(\ref{eq:q0}) equals to
\begin{align}\label{case1forq0}
& \alpha_a  + 1 + \frac{r_a - (r_1 +r_2)}{d} + (r_1 +d-1-r_p)/d \nonumber\\
& = 
\begin{cases}
    \alpha_a + 1 & \text{if } (r_1 \le r_p) \vee (r_a < r_1)\\
    \alpha_a + 2 & \text{if } r_p < r_1 \le r_a 
\end{cases}
\end{align}
Recall that $q=0$ means $\tash{a}_1 \le a$ and $r_1 = \tash{a}_1 \mod d$.
Therefore, 
\begin{align*}
    &\Pr[(q=0) \wedge ((r_1 \le r_p) \vee (r < r_1))] \\
    &= \frac{(d-r_a+r_p)\alpha_a + r_p+1}{p}
\end{align*}
and 
\begin{align*}
    \Pr[(q=0) \wedge (r_p < r_1 \le r_a)] \\
    = \frac{(r_a-r_p)(\alpha_a+1)}{p}
\end{align*}

Next, we consider the $q=1$ case, \ie $a < \tash{a}_1$.
If $r_a+r_p < d$, Eq.~($\ref{eq:q1}$) equals to 
\begin{align}\label{case1forq1}
& \alpha_a  + \frac{r_a + r_p - (r_1 +r_2)}{d} + (r_1-r_p +d-1)/d \nonumber\\
& = 
    \begin{cases}
    \alpha_a  & \text{if } (r_a+r_p < r_1) \vee (r_1 \le r_p) \\
    \alpha_a + 1 & \text{if } (r_p< r_1 \le r_a + r_p).
    \end{cases}
\end{align}

Therefore, if $r_p < r$,
\begin{align*}
    &\Pr[(q=1) \wedge ((r_a + r_p < r_1) \vee (r_1 \le r_p)] \\
    &= \frac{(\alpha_p-\alpha_a)(d-r)-1}{p}
\end{align*}
and
\begin{align*}
    &\Pr[(q=1) \wedge (r_p < r_1 \le r_a+r_p)] \\
    &= \frac{(\alpha_p-\alpha_a-1)r_a+r_p}{p}
\end{align*}

In summary, if $r_p<r_a$ and $r_a+r_p<d$,
\begin{align*}
    &\Pr[\text{output is } \alpha_a] \\
    &= \Pr[(q=1) \wedge (r_a+r_p < r_1) \vee (r_1 \le r_p)] \\
    &= \frac{(\alpha_p-\alpha_a)(d-r_a)-1}{p} 
\end{align*}
and 
\begin{align*}
    &\Pr[\text{output is } \alpha_a +1 ] \\
    &= \Pr\left[
        \begin{alignedat}{1}
            &((q=0) \wedge ((r_1 \le r_p) \vee (r_a < r_1))) \\
            &\vee ((q=1) \wedge (r_p < r_1 \le r_a+r_p)) \\
        \end{alignedat}
        \right]\\
    &= \frac{(d-r_a+r_p)\alpha_a + r_p+1}{p} + \frac{(\alpha_p-\alpha_a-1)r_a+r_p}{p}\\
    &= \frac{(d-2r_a+r_p)\alpha_a + 2r_p+(\alpha_p-1)r_a+1}{p}
\end{align*}
and 
\begin{align*}
    &\Pr[\text{output is } \alpha_a +2 ] = \Pr[(q=0) \wedge (r_p < r_1 \le r)] \\
    &= \frac{(r_a-r_p)(\alpha_a+1)}{p}
\end{align*}

\subsection*{Case 2: $r_p<r_a$ and $d \le r_a+r_p$}
The $q=0$ case is the same as the case 1.

If $d \le r_a+r_p$, Eq.~($\ref{eq:q1}$) equals to 
\begin{align*}
& \alpha_a  + \frac{r_a + r_p - (r_1 +r_2)}{d} + (r_1-r_p +d-1)/d \nonumber\\
& = 
    \begin{cases}
    \alpha_a  & \text{if } (r_a+r_p-d < r_1 \le r_p) \\
    \alpha_a + 1 & \text{if } (r_p< r_1) \vee (r_1 \le r_a+ r_p-d).
    \end{cases}
\end{align*}

If $r_p < r_a$,
\begin{align*}
    &\Pr[(q=1) \wedge (r_a+r_p-d < r_1 \le r_p)] \\
    &= \frac{(\alpha_p-\alpha_a)(d-r_a) -1}{p}
\end{align*}
and
\begin{align*}
    &\Pr[(q=1) \wedge ((r_p< r_1) \vee (r_1 \le r_a+ r_p-d))] \\
    &= \frac{(\alpha_p-\alpha_a-1)r_a+r_p}{p}
\end{align*}

In summary, if $r_p<r_a$ and $d \le r_a+r_p$,
\begin{align*}
    &\Pr[\text{output is } \alpha_a] \\
    &= \Pr[(q=1) \wedge (r_a+r_p-d < r_1 \le r_p)] \\
    &= \frac{(\alpha_p-\alpha_a)(d-r_a)-1}{p}
\end{align*}
and 
\begin{align*}
    &\Pr[\text{output is } \alpha_a +1 ] \\
    &= \Pr\left[
        \begin{alignedat}{1}
            &((q=0) \wedge ((r_1 \le r_p) \vee (r < r_1))) \\
            &\vee ((q=1)\wedge ((r_p< r_1) \vee (r_1 \le r_a+ r_p-d)))
        \end{alignedat}
        \right]\\
    &= \frac{(d-(r_a-r_p))\alpha_a + r_p+1}{p} + \frac{(\alpha_p-\alpha_a-1)r_a+r_p}{p}\\
    &= \frac{(d-2r_a+r_p)\alpha_a + 2r_p+(\alpha_p-1)r_a+1}{p}
\end{align*}
and 
\begin{align*}
    &\Pr[\text{output is } \alpha_a +2 ] = \Pr[(q=0) \wedge (r_p < r_1 \le r_a)] \\
    &= \frac{(r_a-r_p)(\alpha_a+1)}{p}
\end{align*}

The above shows the probability of cases $1$ and $2$ are the same.

\subsection*{Case 3: $r_a\le r_p$ and $d \le r_a+r_p$}

If $r_a \le r_p$, Equation \ref{eq:q0} equals to
\begin{align*}
& \alpha_a  + 1 + \frac{r_a - (r_1 +r_2)}{d} + (r_1 +d-1-r_p)/d \nonumber\\
& = 
\begin{cases}
    \alpha_a & \text{if } r_a < r_1 \le r_p \\
    \alpha_a +1 & \text{if } (r_1 \le r_a) \vee (r_p < r_1)
\end{cases}
\end{align*}
Similar to the previous case, we have
\begin{align*}
    \Pr[(q=0) \wedge (r_a < r_1 \le r_p)] 
    = \frac{(r_p-r_a)\alpha_a}{p}
\end{align*}
and 
\begin{align*}
    &\Pr[(q=0) \wedge ((r_1 \le r_a) \vee (r_p < r_1))] \\
    &= \frac{(d-r_p+r_a)\alpha_a + r_a+1}{p}
\end{align*}

The $q=1$ case, if $r_a \le r_p$,
\begin{align*}
    &\Pr[(q=1) \wedge ((r_a + r_p < r_1) \vee (r_1 \le r_p)] \\
    &= \frac{(\alpha_p-\alpha_a)(d-r_a)-1+r_p-r_a}{p}
\end{align*}
and
\begin{align*}
    \Pr[(q=1) \wedge (r_p < r_1 \le r_a+r_p)] 
    = \frac{(\alpha_p-\alpha_a)r_a}{p}
\end{align*}

In summary,
if $r_a \le r_p$ and $r_a+r_p<d$,
\begin{align*}
    &\Pr[\text{output is } \alpha_a] \\
    &= \Pr\left[
        \begin{alignedat}{1}
            &((q=0) \wedge (r < r_1 \le r_p)) \\
            &\wedge ((q=1) \wedge (r_a+r_p < r_1) \vee (r_1 \le r_p))
        \end{alignedat}
        \right] \\
    &= \frac{(r_p-r)\alpha_a}{p} + \frac{(\alpha_p-\alpha_a)(d-r)-1 +r_p-r}{p} \\
    &= \frac{(\alpha_p-\alpha_a)d+r_p\alpha_a -\alpha_pr_a-1+r_p-r}{p} \\
    &= \frac{p-a + r_p\alpha_a -\alpha_pr_a-1}{p}
\end{align*}
and 
\begin{align*}
    &\Pr[\text{output is } \alpha_a +1 ] \\
    &= \Pr\left[
        \begin{alignedat}{1}
            &((q=0) \wedge ((r_1 \le r_a) \vee (r_p < r_1))) \\
            &\vee ((q=1) \wedge (r_p < r_1 \le r_a+r_p))] 
        \end{alignedat}
        \right]\\
    &= \frac{(d-r_p+r_a)\alpha_a + r_a+1}{p} + \frac{(\alpha_p-\alpha_a)r_a}{p}\\
    &= \frac{(d-r_p)\alpha_a + (\alpha_p+1)r_a+ 1}{p}\\
    &= \frac{a -r_p\alpha_a + r_a\alpha_p  + 1}{p}
\end{align*}

\subsection*{Case 4: $r_a \le r_p$ and $d \le r_a+r_p$}

The $q=0$ case is the same as the case 3.
If $r_a \le r_p$,
\begin{align*}
    &\Pr[(q=1) \wedge (r_a+r_p-d < r_1 \le r_p)] \\
    &= \frac{(\alpha_p-\alpha_a)(d-r_a)-1+r_p-r_a}{p}
\end{align*}
and
\begin{align*}
    &\Pr[(q=1) \wedge ((r_p< r_1) \vee (r_1 \le r_a+ r_p-d))] \\
    &= \frac{(\alpha_p-\alpha_a)r_a}{p}
\end{align*}

In summary, the probability is the same as that in the case 3.

\section{Division Protocol Secure against an Active Adversary with Abort}\label{sect:divisionmal}
We show the division protocol secure against an active adversary with abort in Protocol~\ref{alg:divisionformore}. 
In addition to canceling $q\alpha_p$ out, we adjust the output to make the difference 
between $\frac{a}{d}$ and the output small by adding constants.
Experimental analysis of the output of this protocol is provided in Sect.~\ref{sec:experiment},
which shows on average the relative error is about $0.5$.
Precise analysis of the output distribution will be provided in the full version.
\begin{algorithm}[H] 
\caption{Actively Secure Division by Public Value}
 \label{alg:divisionformore}
 \begin{algorithmic}[1]
  \Functionality $\pash{c}\gets \Div_{(2,3)}^{\textsf{Mal}}(\pash{a}, d)$
  \Require $\pash{a}$ and $d$, where $a$ and $d$ are multiples of $4$
  \Ensure $\pash{c}$, where $c \approx \frac{a}{d}$
    \State Let $\alpha_p$ and $r_p$ be $p=\alpha_p d + r_p$, where $0 \le r_p < d$.
  \State $\pash{q} \gets \Fqt(\pash{a})$
  \State  $z := \begin{cases}
  1 &  \text{if }r_p \ge \intdiv{d}{2}\\
  0 & \text{otherwise}
    \end{cases}$
\State Let $a_i$ be a sub-share of $\pash{a}$, \ie $a_1 + a_2 + a_3 = a \mod p$
\For{$1 \le j \le 3$}
\State $P_j$ and $P_{j+1}$ compute
  $b_j := 
  \begin{cases}
   a_j + (d-r_p) + \intdiv{(d-r_p)}{2} \text{ in }\NN& \text{if } j = 0\\
   a_j & \text{otherwise}
  \end{cases}$
  \State $P_j$ and $P_{j+1}$ set
  $b^\prime_j :=
    \begin{cases}
   \intdiv{b_j}{d} + 1  & \text{if } \frac{b_j}{d}-\intdiv{b_j}{d} \ge \frac{d}{2}\\
   \intdiv{b_j}{d} & \text{otherwise}
  \end{cases}
  $
\EndFor
  \State $\pash{b'}_i := (b'_i, b'_{i+1})$ for $i=1,2,3$
  \State Output $\pash{b^\prime}-(\alpha_p+z)\pash{q}-1 $
 \end{algorithmic}
\end{algorithm}

\section{Extension to Signed Values}\label{seq:signed}

Some of our protocols are only suitable for computation over unsigned values.
To extend the domain of these to include signed values, we make use of a functionality that extracts the sign and the absolute value of a share.
This allows us to do computation over the absolute value, and later adjust the result according to the sign.
We denote by $\Extsignabs$ the function that extracts the sign and absolute value of the input, and functionality $\Fextsignabs$  

\begin{thm} 
The protocol in Figure~\ref{alg:extsignabs} securely implements $\Fextsignabs$ in the ($\Fbitdecomp, \Fmodconv$)-hybrid model in the presence of a passive adversaries. 
\end{thm}

\begin{algorithm}[H] 
\caption{Extract Sign and Absolute for Signed Integer}
 \label{alg:extsignabs}
 \begin{algorithmic}[1]
  \Functionality $(\pash{f},\pash{b}) \gets \Extsignabs(\pash{a})$
  \Require $\pash{a}$
  \Ensure $(\pash{f},\pash{b})$, where $f=1$ if $0 \le a$ and $f=-1$ otherwise, and the least $\ell$ bits of $b$ is $\abs{a}$.
  \Parameter $\ell$, where $\ell$ is the bit-length of $a$, \ie $- 2^{\ell} < a < 2^{\ell}$.
  \State $\pash{a'} = 2^{\ell+1} + \pash{a}$
  \State $(\bash{a_{\ell+1}},\ldots,\bash{a_{1}}) \gets \Fbitdecomp(\pash{a})$.
  \State $\pash{a_{\ell+1}}\gets \Fmodconv(\bash{a_{\ell+1}})$. \Comment{$a_{\ell+1}=1$ if $a$ is positive and $a_{\ell+1}=0$ otherwise}
  \State $\pash{f} := 2\pash{a_{\ell+1}}-1$ \Comment{$f=1$ if $a$ is positive and $f= -1$ otherwise}
  \State $\pash{b} := (1-\pash{a_{\ell+1}})2^{\ell+1} + \pash{f}\cdot\pash{a'}$
  \State Output $(\pash{f},\pash{b})$
 \end{algorithmic}
\end{algorithm}





\section{Bit composition protocol}
\label{sec:bitcomp}

We will make use of functionalities for bit decomposition and composition of shared values, denoted $\Fbitdecomp$ and $\Fbitcomp$, respectively. The former was discussed in Sect.~\ref{sec:prelims}. 
The known bit composition protocol \cite{CCS:ABFKLO18} works in only mod $2^n$,
we, therefore, propose the bit composition protocol in mod $p$.

In the bit-composition protocol,
we want to obtain the composed value $\pash{a}$ on input of its 
binary representation $\bash{a_1},...,\bash{a_\ell}$
by computing addition as $\sum_{i}2^{i-1}\bash{a_i}$.
However, this computation does not work since each $\bash{a_i}$ has the quotient $q_i$
so $2^{i-1}(2q_i+a_i) = 2^{i}q_i + 2^{i-1}a_i$ is added at the $i$-th bit.
The known bit-composition protocol in \cite{CCS:ABFKLO18} cancel $2^iq_i$ out 
by computing $q_i$ recursively for $i\le \ell-1$, and 
they does not need to obtain $q_i$ for $\ell < i$;
their underlying secret sharing is on modulus $2^\ell$ so $2^\ell q_i$ will be 0.
However, this is not the case in modulus $p$
so we have to yield another approach to cancel $2^{\ell} a_i$ out.

In our protocol, we resolve this problem by using the modulus conversion protocol.
By converting shares from modulus 2 into $p$,
$2q_i+a_i$ is changed into $pq_i + a_i$, which is $a_i$ in modulus $p$.
The protocol we use to implement the latter is shown in Figure~\ref{alg:bitdecomposition}.
The round complexity of this protocol is $\ell+1$
if we instantiate $\Fqt$ and $\Fmodconv$ by Protocol~\ref{alg:qt3}
and that in \cite{ACISP:KIMHC18}, where $\ell$ is the maximum bit-length of shared secret.

\begin{algorithm}[t]\small 
\caption{Bit Composition}
 \label{alg:bitdecomposition}
 \begin{algorithmic}[1]
  \Functionality $\pash{\displaystyle\sum_{i<\ell}2^ia_i}\gets \BC(\bash{a_1},\dots,\bash{a_{\ell}})$
  \Require $\bash{a_1},\dots,\bash{a_{\ell}}$
  \Ensure $\pash{\displaystyle\sum_{i=1}^{\ell}2^i a_i}$
  \Parameter The bit-length of secret $\ell$ 
  \State Each $P_i$ sets $\bash{b_1}_i:=[0]$  \Comment{Set its sub-share as $(0,0)$}
  \State $\bash{a_1^\prime} := \bash{a_1}$
  \State $\bash{q_1'}\gets\Fqt(\bash{a_1^\prime})$
  \For{$i = 2 \text{ to } \ell$}
  \State $\bash{a_i^\prime} := \bash{a_i} - \bash{q_{i-1}^\prime} -\bash{b_{i-1}}$
  \State $\bash{b_i} := ((1-\bash{a_i}) + \bash{b_{i-1}}) \cdot (\bash{q_{i-1}} + \bash{b_{i-1}}) + \bash{b_{i-1}}$ \Comment{$\bash{b_i}$ is the borrow of $i$-th bits}
  \State $\bash{q_i^\prime}\gets\Fqt(\bash{a_i^\prime})$
  \EndFor
  \State $\pash{b_{\ell}} \gets \Fmodconv(\bash{b_{\ell}} )$
  \State $\pash{q_{\ell}^\prime} \gets \Fmodconv(\bash{q_{\ell}^\prime})$
  \State $\pash{a^\prime}_j \gets \displaystyle\sum_{i\le\ell}2^{i-1}\bash{a_i^\prime}_j \mod p$
  \State Output $\pash{a^\prime}-2^\ell(\pash{b_{\ell}}+\pash{q_{\ell}^\prime})$
 \end{algorithmic}
\end{algorithm}

\begin{thm} 
The protocol in Figure~\ref{alg:bitdecomposition} securely implements $\Fbitcomp$ in the ($\Fqt$, $\Fmult$, $\Fmodconv$)-hybrid model in the presence of a passive adversary. 
\end{thm}

\section{Details of Secure Deep Neural Network}\label{sec:defernn}

\subsection{Notation}
$L$ denotes the layer number, and when the number of hidden layers is $n$, 
the input layer is $L=0$ and the output layer is $L=n+1$.
The value $d_{L=j}$ denotes the number of neurons in layer $j$, and $N_j^i$ denotes the $i$-th neuron in layer $L=j$. 
The strength of the coupling of neurons is described by parameters $w_i$. 
Learning is a process that (iteratively) updates the parameters to obtain the appropriate output.

In this section, the unit of processing is a matrix,
so we use different notation.
Let $A = (a_{i,j})$ denote a matrix,
$A \pm B$ and $A \cdot B$ denote the matrix addition/subtraction and product,
and $A \circ B$ denote the element-wise multiplication.
When we apply an algorithm $\textsf{Func}$ with each element $a_{i,j}$ in a matrix $A$,
we describe it as $\textsf{Func}(A)$, such as $\sqrt{A}$ and $\Fbitdecomp(A)$.

\subsection{Details of Adam}

We introduce the main process used in Adam.
The process is the same in each layer so we omit the layer index. A variable $t$ indicates the iteration number of the learning process,
\eg the value $G_t$ denotes the gradient of the $t$-th iteration. 
In addition, $M,V,\hat{M}, \hat{V}$ are matrices of the same size as $G$, and $M$ and $V$ are initialized by 0. 
Here the superscript $t$, such as $\beta^{t}$, represents the $t$-th power of $\beta$. 
Adam proceeds as follows.
\begin{align}\small
    M_{t+1} &= \beta_1 M_t + (1-\beta_1)G_t \nonumber\\ 
    V_{t+1} &= \beta_2 V_t + (1-\beta_2)G_t\circ G_t \nonumber\\
    \hat{M}_{t+1} &= \frac{1}{1-\beta_1^t}M_{t+1} \nonumber\\
    \hat{V}_{t+1} &= \frac{1}{1-\beta_2^t}V_{t+1} \nonumber\\
    W_{t+1} &= W_t - \frac{\eta}{\sqrt{\hat{V}_{t+1}}+\epsilon}\circ \hat{M}_{t+1} \label{eq:adam2}
\end{align}
where $\circ$ denotes the element-wise multiplication of matrices.

\subsection{Secure Protocols for Neural Network}

\subsubsection{ReLU and ReLU$^\prime$ functions}
The ReLU$^\prime$ function extracts the sign of the input as a bit value $b \in \bit$.
Hence, we can use the same approach as in Protocol~\ref{alg:extsignabs} to implement this.
The ReLU function can obtained by simply multiplying the input with the output of ReLU$^\prime$.
We show secure protocols for the ReLU$^\prime$ and ReLU functions in Protocol~\ref{alg:ReLUprime} and \ref{alg:ReLU}.

\begin{algorithm}[htbp] \small
\caption{Secure ReLU$^\prime$ Function}
 \label{alg:ReLUprime}
 \begin{algorithmic}[1]
  \Functionality 
  $\pash{Z} \gets \ReLU'(\pash{U})$
  \Require A matrix $\pash{U}$
  \Ensure $\pash{Z}$, where each element $z_{i,j}$ in $Z$ is 0 if $z_{i,j} \le 0$ and 1 otherwise
  \Parameter $\ell$, where each element in $Z$ is between $2^{-\ell}+1$ and $2^\ell-1$.
    \State $\pash{U'} = 2^{\ell+1} + \pash{U}$ \Comment{Add $2^{\ell+1}$ to each element}
  \State $(\bash{U^{(\ell+1)}},\ldots,\bash{U^{(1)}}) \gets \Fbitdecomp(\pash{U})$ 
  \State $\pash{Z} \gets \Fmodconv(\bash{U^{(\ell+1)}})$. \Comment{Each element in $U^{(\ell+1)}$ is 1 if the corresponding element in $U$ is positive and 0 otherwise}
  \State Output $\pash{Z}$
 \end{algorithmic}
\end{algorithm}

\begin{algorithm}[htbp] \small
\caption{Secure ReLU Function}
 \label{alg:ReLU}
 \begin{algorithmic}[1]
  \Functionality 
  $\pash{Y} \gets \ReLU(\pash{U})$
  \Require A matrix $\pash{U}$
  \Ensure $\pash{Y}$, where $Y \gets \textbf{ReLU}(U)$.
  \Parameter $\ell$, where each element in $U$ is between $2^{-\ell}+1$ and $2^\ell-1$.
    \State $\pash{Z} = \ReLU'(\pash{U})$
    \State $\pash{Y} \gets \pash{Z} \circ \pash{U}$
 \end{algorithmic}
\end{algorithm}

\subsubsection{Softmax function}
As shown in Eq.~\ref{eq:softmax},
the softmax function is computed via exponential and inversion functions,
which we can implement via Protocol~\ref{alg:exponent} and \ref{alg:inversion}, respectively.
We show a secure protocol for the softmax function in Protocol~\ref{alg:softmax}.

\begin{algorithm}[htbp] \small
\caption{Secure Softmax Function}
 \label{alg:softmax}
 \begin{algorithmic}[1]
  \Functionality 
  $\pash{Y} \gets \softmax(\pash{U})$
  \Require A matrix $\pash{U}=(\pash{u_{1,1}}, \ldots, \pash{u_{m,n}})$
  \Ensure $\pash{Y}$, where $Y \gets \textbf{softmax}(U)$.
    \State $\pash{\vec{u_i}} := (\pash{u_{i,1}}, \ldots, \pash{u_{i,n}})$
    \For{$i = 1$ to $m$ (in parallel)}
        \For{$j = 1$ to $n$ (in parallel)}
    \State Set $\pash{\vec{u}_{i,j}} = (\pash{u_{i,j}}, \ldots, \pash{u_{i,j}})$ of length $n$ \Comment{all the elements are the same}.
    \State $\pash{\vec{b}_i} \gets \pash{\vec{u}_i}-\pash{\vec{u}_{i,j}}$
    \State $\pash{\vec{c}_i} \gets \Exponent(\pash{\vec{b}_i})$ \Comment{$\pash{\vec{c}_i}$ is $(\pash{e^{u_{i,1}-u_{i,j}}}, \ldots, \pash{e^{u_{i,n}-u_{i,j}}})$}
    \State $\pash{\sum_{k=1}^{n}e^{u_{i,k}-u_{i,j}}} := \sum_{k=1}^{n}\pash{c_{i,k}}$
        \Comment{Sum all the elements in $C$}
     \State $\pash{y_{i,j}} \gets \Inv(\pash{\sum_{k=1}^{n}e^{u_{i,k}-u_{i,j}}})$
    \EndFor
    \EndFor
    \State $\pash{Y} := (\pash{y_{1,1}}, \dots, \pash{y_{m,n}})$
 \end{algorithmic}
\end{algorithm}

\subsubsection{Simultaneous offset management and right shift}
The deep neural network computations are done using fixed-point arithmetic,
and hence require truncation of the lower bits after each multiplication.
At the same time, in the training process,
we have to divide intermediate values with the batch size $m$, and 
additionally, results are scaled with the learning rate $\eta$, which is typically small \eg $0.001$.

This opens up the possibility of an optimization in which the above operations, where appropriate, can be performed simultaneously using a single  arithmetic right-shift.
In particular, setting the batch size as a multiple of 2, such as $m=2^7=128$,
and using an approximate learning rate $\eta' :=  2^{-10} \approx 0.001$, makes this optimization very simple to implement.
Note that all of our protocols are designed to allow the input and output offsets to be chosen freely, and adding optimization such as the above, by adjusting the offsets, is trivial.
In the following, we use the notation $\cdot_H$ to denote (matrix) multiplication with an additional right-shift $H$, and $\InvSqrt(\cdot, \nu)$ to denote computation of the inverse square root with an additional right-shift $\nu$. 

\subsubsection{Secure Feedforward Deep Neural Networks with Adam}
Let $n$ be the number of hidden layers,
$\widehat{\beta}_{1,t} = \frac{1}{1-\beta_1^t}$,
and $\widehat{\beta}_{1,t} = \frac{1}{1-\beta_2^t}$.
We show a protocol for secure feedforward deep neural networks with Adam in Protcool~\ref{alg:NNAdam}.
In the protocol description, for brevity, we omit the division for signed and unsigned integers required by fixed-point multiplication.
This should be done following the inner product and element-wise multiplication.


\begin{algorithm}[htbp] \small
\caption{Secure Feedforward Neural Network with Adam}
 \label{alg:NNAdam}
 \begin{algorithmic}[1]
  \Functionality 
  $\pash{W^l} \gets \textsf{FFNN\_Adam}(\pash{X},\pash{T}, \pash{W^0},\ldots, \allowbreak\pash{W^n}, \pash{M^0},\dots,\pash{M^n},\pash{V^0},\dots,\pash{V^n})$
  \Require Features of trained data $\pash{X}$, their label $\pash{T}$, initialized parameters $\pash{W^l}$, and  vectors initialized by 0 $\pash{M^l}$ and $\pash{V^l}$ for $0 \le l \le n$
  \Ensure Updated parameters $\pash{W^l}$ for $0 \le l \le n$
  \Parameter $\eta',\beta_1,\beta_2,\hat{\beta}_{1,t}, \hat{\beta}_{2,t}$
  \State ---Forward propagation---
  \State $\pash{U^1} \gets \pash{W^0} \cdot \pash{X}$
  \State $\pash{Y^1} \gets \ReLU(\pash{U^1})$
  \For{$i=1$ to $n-1$}
    \State $\pash{U^{i+1}} \gets \pash{W^i} \cdot \pash{Y^i}$
    \State $\pash{Y^{i+1}} \gets \ReLU(\pash{U^{i+1}})$
  \EndFor
  \State $\pash{U^{n+1}} \gets \pash{W^n} \cdot \pash{Y^n}$
  \State $\pash{Y^{n+1}} \gets \softmax(\pash{U^{n+1}})$
  \State ---Back propagation---
  \State $\pash{Z^{n+1}} \gets \pash{Y^{n+1}} - \pash{T}$
  \State $\pash{Z^{n}} \gets \ReLU'(\pash{U^{n}})\circ (\pash{Z^{n+1}} \cdot \pash{W^n})$
  \For{$i=1$ to $n-1$}
    \State $\pash{Z^{n-i}} \gets \ReLU'(\pash{U^{n-i}})\circ(\pash{Z^{n-i+1}}\cdot \pash{W^{n-i}})$
  \EndFor
  \State ---Gradient evaluation---
  \State $\pash{G^{0}} \gets \pash{Z^{1}} \cdot_{H} \pash{T}$ 
  \For{$i=1$ to $n-1$}
    \State $\pash{G^{i}} \gets \pash{Z^{i+1}} \cdot_{H} \pash{Y^i}$
  \EndFor
  \State $\pash{G^{n}} \gets \pash{Z^{n+1}} \cdot_{H} \pash{Y^n}$
  \State ---Parameter update by Adam---
  \For{$i=0$ to $n$}
  \State $\pash{M^i} \gets \beta_1\pash{M^i} + (1-\beta_1)\pash{G^i}$
  \State $\pash{V^i} \gets \beta_2\pash{V^i} + (1-\beta_2)\pash{G^i} \circ \pash{G^i}$
  \State $\pash{\widehat{M}^i} \gets \widehat{\beta}_{1,t}\pash{M^i}$
    \State $\pash{\widehat{V}^i} \gets \widehat{\beta}_{2,t}\pash{V^i}$
    \State $\pash{\widehat{G}^i} \gets \InvSqrt(\pash{\widehat{V}^i},\eta')$ 
    \State $\pash{\widehat{G}^i} \gets \pash{\widehat{G}^i} \circ \pash{\widehat{M}^i}$
    \State $\pash{W^i} \gets \pash{{W}^i} - \pash{\widehat{G}^i}$
    \EndFor
 \end{algorithmic}
\end{algorithm}

\subsubsection{Extension to Convolutional Deep Neural Networks}
\newcommand{\MaxFlag}{\mathsf{MaxFlag}}

In order to extend the feedforward neural network to the covolutional neural network, we need to implement a convolutional layer, batch normalization, and max-pooling. The convolutional layer can be computed as the usual fully connected layer, and batch-normalization can be computed trivially as long as the inverse of square root can be computed, since the rest is a linear operation. In the max-pooling, we need the maximum value in a vector and a flag indicating location of the maximum value to compute forward and backward propagation. We obtain those by repeatedly compare values in the vector. 

The secure max-pooling is shown in Protocol~\ref{alg:maxpool}. 
This protocol is defined as a recursive function.
We define several functionalities and notations for this protocol:
Let $\Fcompare$ be the functionality on input $\pash{\vec{a}}=(\pash{a_1},\ldots, \pash{a_n})$ and $\pash{\vec{b}}=(\pash{b_1},\ldots, \pash{b_n})$
outputs $\pash{\vec{c}}$, where $c_i = 1$ if $a_i \ge b_i$ and $0$ otherwise for $1\le i \le n$. A parallel execution of a comparison protocol, e.g., \cite{ACISP:KIMHC18}, can realize $\Fcompare$.
Let $\Fcondassignshare$ be the functionality that is similar to $\Fcondassign$, but assigned values are not plaintexts but shares: on input of $(\pash{\vec{b}},\pash{\vec{a}},\pash{\vec{c}})$ such that $c_i \in \bit$, 
outputs $\pash{\vec{d}}$, where $d_i = a_i$ if $c_i =1$ and $d_i=b_i$ otherwise for $1\le i \le n$. This functionality can be realized in a similar way that computes
$\pash{c_i}\cdot\pash{a_i} + (1-\pash{c_i})\pash{b_i}$ using multiplication protocol in parallel.
For two vectors $\pash{\vec{e}}=(\pash{e_1},\ldots,\pash{e_n})$ and
$\pash{\vec{f}}=(\pash{f_1},\ldots,\pash{f_m})$, $\pash{\vec{e}}||\pash{\vec{f}}$ denote the concatenation of the vectors, i.e., $\pash{\vec{e}}||\pash{\vec{f}} = (\pash{e_1},\ldots,\pash{e_n},\pash{f_1},\ldots,\pash{f_m})$.

\begin{algorithm}[htbp] \small
\caption{Secure max-pooling}
 \label{alg:maxpool}
 \begin{algorithmic}[1]
  \Functionality 
  $(\pash{y}, \pash{\overrightarrow{f}}) \gets \MaxFlag(\pash{\overrightarrow{x}})$
  \Require A vector of length $n$, $\pash{\overrightarrow{x}} = (\pash{x_1}, \ldots ,\pash{x_n})$
  \Ensure The maximum $\pash{y}$ and its original location $\pash{\overrightarrow{f}} = (\pash{f_1}, \ldots ,\pash{f_n})$, where $y = \max_{1\le i \le n}x_i$ and $f_j = 1$ if $y=x_j$ and $f_j=0$ otherwise for $1\le j \le n$.\footnote{If there are multiple maximum values, only the location of one of them will be set to 1, and the others will be set to 0.}
    \If{$n=1$}
        \Return $\pash{x_1},(\pash{1})$.
    \Else
        \State $m=n/2$
        \State $\pash{\vec{a}} := (\pash{x_1},\ldots,\pash{x_{m}})$
        \State $\pash{\vec{b}} := (\pash{x_{m+1}},\ldots,\pash{x_{n}})$
        \State $\pash{\vec{c}} \gets \Fcompare(\pash{\vec{a}},\pash{\vec{b}})$
        \State $\pash{\vec{d}} \gets \Fcondassignshare(\pash{\vec{b}},\pash{\vec{a}},\pash{\vec{c}})$
        \If{$n \mod 2 = 0$}
            \State $(\pash{y}, \pash{\vec{e}}) \gets \MaxFlag(\pash{\vec{d}})$
            \State $\pash{\vec{z}} \gets \Fmult(\pash{\vec{e}},\pash{\vec{c}})$ 
            \State $\pash{\vec{f}} := \pash{\vec{z}} || (\pash{\vec{e}} - \pash{\vec{z}})$
        \Else
            \State $(\pash{y}, \pash{\vec{e}} || \pash{r}) \gets \MaxFlag(\pash{\vec{d}}||\pash{x_n})$
            \State $\pash{\vec{z}} \gets \Fmult(\pash{\vec{e}},\pash{\vec{c}})$ 
            \State $\pash{\vec{f}} := \pash{\vec{z}} || (\pash{\vec{e}} - \pash{\vec{z}}) || \pash{r}$
        \EndIf
     \EndIf
 \end{algorithmic}
\end{algorithm}

\subsection{Possibilities of other ML-related algorithms}
This section mentions some ML-related algorithms that can be implemented in secure computation, other than those mentioned in the text.

Shortcut \cite{ShortcutResnet} is a technique to add input not only to the next layer but also to a latter layer. This technique is easy to implement in secure computation by adding the input to the later layer.

Dropout \cite{Dropout} is a method of dropping the information from random neurons to prevent overlearning. This method can also be implemented in secure computation. If it is public which neurons will be dropped, the parties generate a random number and set the weight of the neuron, corresponding to the random number, to 0. If we want to hide the location of a neuron to be dropped, the parties generate $\pash{0}$ or $\pash{1}$ for each neuron, shuffle them \cite{LWZ11}, and then multiply them by the weight.

Although we implemented Adam in this paper, other optimization methods, such as Adabound \cite{Adabound} and RAdam \cite{RAdam}, consist of square root, reciprocal, and division. 
Therefore, it should be possible to implement these by using our seamless paradigm. 
This is a future work to implement them and compare their efficiency in the context of secure computation.

\section{Accuracy and Throughput of Division and Elementary Functions}\label{sec:expdivelementary}

\subsection{Accuracy}
We compare the output of our secure division and other elementary protocols
with the \emph{real-valued} function \emph{in the clear}.

\smallstart{Measuring Error in Division.}
We first give our experimental results on division and truncation, summarized in Table~\ref{tab:accdivision}.
We measured the error of the L1-norm as follows. We use inputs from $1$ to $n$, where 
$n=10,000$. 
The average-case and worst-case error refers to the value 
$\frac{1}{n}\sum_i\left| \frac{a_i-c_i}{c_i} \right|$ 
and $\max_i\left| \frac{{a_i}-c_i}{c_i} \right|$, respectively,
where $c_i$ is the correct output and $a_i$ is the output of our protocol.
From the table, we can observe that 
the output of our protocol is close to the real-valued division.
For example, the L1 error of passively secure division protocol is $0.335$ on average,
meaning that the actual error is $\frac{0.335}{2^t}$ if the input offset is $t$.

 \begin{table}[htbp]\footnotesize
 \caption{Accuracy of division and truncation} 
 \vspace{-1mm}
 \label{tab:accdivision}
 \centering
 \begin{tabular}{crrrr} 
 \toprule
 & \multicolumn{2}{c}{Passive} & \multicolumn{2}{c}{Active}\\ 
  & Average & Worst & Average & Worst \\
  \hline  
 Truncation & 0.3304 & 1.000 & 0.483 & 1.875 \\ 
 Division & 0.335 & 1.059 & 0.495 & 2.000 \\
 \bottomrule
 \end{tabular}
 \end{table}

\smallstart{Measuring Error in Elementary Protocols.}
We summarize the experimental results for error in elementary protocols in Table~\ref{tab:accelementary}. Again, we use inputs from $1$ to $n$, where 
$n=10,000$. We set the offset as $\ell=10$.
We set the divisor in the division with private divisor as $3$, with offset being $0$.
Average-case and worst-case accuracy number refers to the value
$-\log \left(\frac{1}{n}\sum_i\left| \frac{a_i-c_i}{c_i} \right|\right)$ 
and $-\log \left( \max_i\left| \frac{{a_i}-c_i}{c_i} \right| \right)$, respectively,
where $c_i$ is the correct output and $a_i$ is the output of our protocol.
This accuracy number $x$ implies that the most significant $x$ bits in the output of the protocols will be equal to those of a corresponding function in the clear.
We prepare the correct output $c_i$ by using functions implemented in the C Language with double precision.

\begin{table}[htbp]\footnotesize
\caption{Accuracy of elementary functions}
\vspace{-1mm}
\label{tab:accelementary}
\centering
\begin{tabular}{crrrr} 
\toprule
 & \multicolumn{2}{c}{Passive} & \multicolumn{2}{c}{Active}\\
 & \!\!\!\!\! Ave [bit] & \!\!\!\!\! Worst [bit] & \!\!\!\!\! Ave [bit] & \!\!\!\!\!Worst [bit]\\ 
\hline  \noalign{\smallskip}
Inversion & 29.62 & 27.27 & 28.97 & 26.77\\
Division (private diviser) & 29.61 & 27.2 & 28.99 & 26.59\\ 
Square root & 29.33 & 27.02 & 28.88 & 26.64\\
Inverse of square root & 29.34 & 27.05 & 28.86 & 26.55\\
Exponential & 25.75 & 24.1 & 25.34 & 23.13\\
\bottomrule
\end{tabular}
\end{table}

In the implementation, we use an internal precision that represents 
how many bits we used to represent intermediate values to obtain high accuracy of the output.
The accuracy also depends on how many terms we compute of the Taylor of Newton series.
In this experiment,
we used 29-bit internal precision and computed by the 4-th term of the Taylor series for the inversion,
28-bit and the 6-th term for the inverse of square root,
25-bit and the 4-th term for exponential functions.
The threshold for our hybrid table-lookup/series-expansion technique for exponential functions is set to $t=4$.

The results show that our protocols achieve more than 23-bit accuracy,
even in the worst case.
Therefore, our protocols of elementary functions 
is as accurate as the Single-precision number,
which also have 23-bit accuracy.

\subsection{Throughput}
We show the throughput of our protocol in Table~\ref{tab:throughputelementary}.
Note that for throughputs, higher is better.
A secret of $29$-bit length is used on this experiment.
We listed the throughput results of state-of-the-art protocols from the latest version of Sharemind~\cite{Ran17} as a reference.\footnote{It is difficult to compare throughputs directly.
They used a different machine and a similar but different modulus, and their throughput is for floating-point arithmetic. 
The last comes from the fact that the throughput of the fixed-point arithmetic is \emph{slower} than that of the floating-point arithmetic in \cite{Ran17}.
}
In the settings where the number of records is large, our implementation is an order of magnitude faster than \cite{Ran17}, while, in the setting with a small number of records, they are comparable.

\begin{table}[htbp]\footnotesize
\caption{Throughput of elementary functions [M op/s]} 
\vspace{-1mm}
\label{tab:throughputelementary}
\centering
\begin{tabular}{ccccc} 
\toprule
&& \multicolumn{2}{c}{Passive} & Active\\
 & & 1,000 [records] & 1,000,000 & 1,000,000 \\ 
\hline  \noalign{\smallskip}
\multirow{4}{*}{\textbf{Ours}} &Truncation & 0.612 & 14.80 & 4.51\\
& Inversion & 0.0219 & 1.179 & 0.236\\ 
& Square root & 0.0147 & 0.608 & 0.136 \\
& Exponential & 0.0355 & 1.017 & 0.237\\
\hline
\multirow{4}{*}{\cite{Ran17}} &
Truncation & 0.833 & 1.82 & - \\
& Inversion & 0.0547 & 0.0567 & -\\ 
& Square root & 0.0559 & 0.0574 &-\\
& Exponential & 0.0350 & 0.0391 &-\\
\bottomrule
\end{tabular}
\end{table}

\section{Additional Experiments with Varying Parameters.}
\label{sec:deferexperiment}
In this section, we describe additional experiments, deferred from Section~\ref{sec:experiment}.
We measured the running time when varying parameters such as the batch size, the number of neurons,
the number of hidden layers, and the number of input records.
These numbers are useful to estimate the running time for different networks and data.

\smallstart{Varying Batch Sizes.}
We show how the running time (for 1 epoch) depends on batch sizes when $n = 2, d_1 = d_2 = 128$ in Table~\ref{tab:numbatchsize}. 
%
Execution time for 1 epoch is almost in inverse proportion to batch size.
Therefore, increasing the batch size helps to implement faster learning.
This likely comes from the fact that the large batch size implies
fewer parameter updates.

\begin{table}[htb]\footnotesize
  \begin{center}
    \caption{Execution Time for each Batch Size}
    \begin{tabular}{rcccc} 
    \toprule
Batch Size & 64 & 128 & 256 & 512\\
\hline
Execution Time [s] & $225$ & $117$ & $64$ & $37$\\
\bottomrule
    \end{tabular}
    \label{tab:numbatchsize}
  \end{center}
\end{table}

\smallstart{Varying the Number of Neurons.}
We show the running time (for 1 epoch) depends on the number of neurons when $n=2, m=128$ in Table~\ref{tab:numneuron}.
Even for a large number of neurons, \ie $256$, the execution time is only $198$ seconds.

\begin{table}[htb]\footnotesize
  \begin{center}
    \caption{Execution Time for each \# of Neurons }
    \begin{tabular}{r|cccc} 
    \toprule
\# of Neurons & 32 & 64 & 128 & 256\\
\hline
Execution Time & 68 & 85 & 117 & 304\\
\bottomrule
    \end{tabular}
    \label{tab:numneuron}
  \end{center}
\end{table}

\smallstart{Varying the Number of Hidden Layers.}
We show how the running time (for 1 epoch) depends on the number of hidden layers when $m=128, d_1=d_2=128$ in Table~\ref{tab:numhiddenlayer}.
For 4 hidden layers, the execution time is less than $3$ minutes.
This result roughly shows that we need approximately 15 seconds more to have another hidden layer. 
This helps us estimate execution time for deeper networks.

\begin{table}[htb]\footnotesize
  \begin{center}
    \caption{Execution Time for each \# of Hidden Layers }
    \begin{tabular}{r|cccc} 
    \toprule
\# of Hidden Layers & 1 & 2 & 3 & 4\\
\hline
Execution Time & 102 & 117 & 133 & 149\\
\bottomrule
    \end{tabular}
    \label{tab:numhiddenlayer}
  \end{center}
\end{table}

\smallstart{Large Data: 100 Features $\times$ 10 Million Data Samples.} 
Table~\ref{tab:100attr} shows the execution time per epoch using 100 attributes $\times$ 10 million data samples.
The parameters $n = 2, d_0 = 100, d_1 = d_2 = 128, d_{3}=10$ are used.

\begin{table}[htb]\footnotesize
\centering
\caption{Execution Time for 100 Attributes and 10 Million Data [s]}
\label{tab:100attr}
  \begin{tabular}{cc}
  \toprule
Batch Size& Our Protocol\\
\hline
$512$& $4,047$ [s] ($1.12$ [h]) \\
$1024$& $2,724$ [s] ($0.76$ [h]) \\
  \bottomrule
  \end{tabular}
\end{table}

\end{document}

%% file: moritastyle.tex
\newcommand{\ie}{\textit{i.e., }}
\newcommand{\eg}{\textit{e.g., }}

\newcommand{\etal}{~\textit{et al.}}
\newcommand{\bit}{\{0, 1\}}
\newcommand{\ReLU}{\textbf{ReLU}}
\newcommand{\softmax}{\textbf{softmax}}

%% file: preamble.tex
\usepackage{underscore}
\usepackage{amsmath}
\usepackage{stmaryrd} 

\usepackage[noend]{algpseudocode}
\usepackage{algorithm}

\algnewcommand\Functionality{\item[\textbf{Functionality:}]}
 \renewcommand{\algorithmicrequire}{\textbf{Input:}}
 \renewcommand{\algorithmicensure}{\textbf{Output:}}
\algnewcommand\Parameter{\item[\textbf{Parameter:}]}
\makeatletter
\renewcommand*{\ALG@name}{Protocol}
\makeatother
\usepackage{caption}
\usepackage{subcaption}

\allowdisplaybreaks[1]

\usepackage[
	n,
	operators,
	advantage,
	sets,
	adversary,
	landau,
	probability,
	notions,	
	logic,
	ff,
	mm,
	primitives,
	events,
	complexity,
	asymptotics,
	keys]{cryptocode}
\usepackage{csquotes}
\usepackage{dashbox}

\usepackage{todonotes}
\usepackage{url}
\usetikzlibrary{shapes.callouts}
\usepackage{listings}
\usepackage{trace}
\usepackage{mdframed}

\usepackage{tabularx,bigdelim}

\newcommand{\pash}[1]{%
\llbracket {#1} \rrbracket%
 }

\newcommand{\bash}[1]{%
[{#1}]
 }

\newcommand{\tash}[1]{%
\langle\!\langle {#1} \rangle\!\rangle%
 }

\newcommand{\texthead}[1]{
\noindent\textbf{\textbf{#1}}
}

\newcommand{\Div}{\mathsf{Div}}

\newcommand{\ptotconvert}{\mathsf{ConvertToAdd}}
\newcommand{\ttopconvert}{\mathsf{ConvertToRep}}
\newcommand{\QT}{\mathsf{QT}}
\newcommand{\MSNZBFit}{\mathsf{MSNZBFit}}
\newcommand{\MSNZBFitExt}{\mathsf{MSNZBFitExt}}
\newcommand{\BC}{\mathsf{BC}}
\newcommand{\Inv}{\mathsf{Inv}}
\newcommand{\Divpriv}{\Div_\mathsf{priv}}

\newcommand{\InvSqrt}{\mathsf{InvSqrt}}
\newcommand{\Sqrt}{\mathsf{Sqrt}}
\newcommand{\Exponent}{\mathsf{Exponent}}

\newcommand{\Extsignabs}{\mathsf{ExtSignAbs}}

\newtheorem{thm}{Theorem}

\AtBeginDocument{%
  \providecommand\BibTeX{{%
    \normalfont B\kern-0.5em{\scshape i\kern-0.25em b}\kern-0.8em\TeX}}}

\newtheorem{myfunctionality}[thm]{FUNCTIONALITY}{\bf}{}   
\newcommand{\functionality}[4]{
\begin{figure*}[#3]
\centering
\framebox[\width][c]{
    \small
    \hbox{\quad
    \begin{varwidth}[c]{0.9\textwidth}
    \begin{myfunctionality}
    [#1]
    \label{#2}
    \end{myfunctionality}
    {#4}
    \end{varwidth}
    \quad}
}
\end{figure*}
}

\newcommand{\intdiv}[2]{
{
{#1}/{#2}
}
}